\documentclass[
    aps,
    prx,
    letterpaper,
    nobalancelastpage,
    twocolumn,
    superscriptaddress,
    nofootinbib,
    longbibliography
]{revtex4-2}

\usepackage{graphicx}
\usepackage{amsmath}
\usepackage{amssymb}
\usepackage[english]{babel}
\usepackage{color}
\usepackage[version=4]{mhchem}
\usepackage[hidelinks]{hyperref}
\usepackage{dsfont}
\usepackage{multirow}
\usepackage{booktabs}
\usepackage{array}

\usepackage{soul}
\newcommand{\abs}[1]{\left| #1 \right|}

\newcommand{\ket}[1]{{| {#1} \rangle}}
\newcommand{\bra}[1]{{\langle {#1} |}}
\newcommand{\braket}[2]{\langle {#1} | {#2} \rangle}
\newcommand{\ketbra}[2]{| {#1} \rangle \langle {#2} |}
\newcommand{\tld}[1]{\widetilde{#1}}
\newcommand{\dd}[1]{\,\t{d}{#1}\,}

\renewcommand{\t}[1]{\text{#1}}

\newcommand{\UIUC}{
    Department of Physics,
    The University of Illinois at Urbana-Champaign,
    Urbana, IL 61801, USA
}
\newcommand{\PUa}{
    Department of Physics and Astronomy,
    Purdue University,
    West Lafayette, IN 47907, USA
}
\newcommand{\PUb}{
    Purdue Quantum Science and Engineering Institute,
    Purdue University,
    West Lafayette, IN 47907, USA
}

\renewcommand{\cite}[1]{\mbox{\citep{#1}}}

\addto\captionsenglish{}
\addto\captionsenglish{\renewcommand\thefigure{\arabic{figure}}}

\begin{document}

\title{Analyzing the Rydberg-based \textit{omg} architecture for $^{171}$Yb nuclear spins}
\author{Neville Chen}
\thanks{These three authors contributed equally}
\affiliation{\UIUC}
\author{Lintao Li}
\thanks{These three authors contributed equally}
\affiliation{\UIUC}
\author{William Huie}
\thanks{These three authors contributed equally}
\affiliation{\UIUC}
\author{Mingkun Zhao}
\affiliation{\UIUC}
\author{Ian Vetter}
\affiliation{\UIUC}
\author{Chris H. Greene}
\affiliation{\PUa}
\affiliation{\PUb}
\author{Jacob P. Covey}\email{jcovey@illinois.edu}
\affiliation{\UIUC}

\begin{abstract}
    Neutral alkaline earth(-like) atoms have recently been employed in atomic
    arrays with individual readout, control, and high-fidelity Rydberg-mediated
    entanglement. This emerging platform offers a wide range of new quantum
    science applications that leverage the unique properties of such atoms:
    ultra-narrow optical ``clock'' transitions and isolated nuclear spins.
    Specifically, these properties offer an optical qubit (\textit{o}) as well
    as ground (\textit{g}) and metastable (\textit{m}) nuclear spin qubits, all
    within a single atom. We consider experimentally realistic control of this
    \textit{omg} architecture and its coupling to Rydberg states for
    entanglement generation, focusing specifically on ytterbium-171
    ($^{171}\t{Yb}$) with nuclear spin $I = 1/2$. We analyze the $S$-series
    Rydberg states of $^{171}\t{Yb}$, described by the three spin-$1/2$
    constituents (two electrons and the nucleus). We confirm that the $F = 3/2$
    manifold -- a unique spin configuration -- is well suited for entangling
    nuclear spin qubits. Further, we analyze the $F = 1/2$ series -- described
    by two overlapping spin configurations -- using a multichannel quantum
    defect theory. We study the multilevel dynamics of the nuclear spin states
    when driving the clock or Rydberg transition with Rabi frequency
    $\Omega_\t{c} = 2 \pi \times 200\,\t{kHz}$ or $\Omega_\t{R} = 2 \pi \times
    6\,\t{MHz}$, respectively, finding that a modest magnetic field ($\approx
    200\,\t{G}$) and feasible laser polarization intensity purity ($\lesssim
    0.99$) are sufficient for gate fidelities exceeding 0.99. We also study single-beam Raman rotations of the nuclear spin qubits and identify a ``magic'' linear polarization angle with respect to the magnetic field at which purely $\sigma_x$ rotations are possible. 
\end{abstract}
\maketitle

\section{Introduction}\label{sec: Intro}
Individually trapped neutral atoms with interactions mediated by highly-excited
Rydberg states have become a prominent platform for quantum
science~\cite{Saffman2010, Browaeys2020, Morgado2021}. Most research to date
with arrays of neutral atoms has been conducted with alkali species, but
alkaline earth(-like) atoms (AEAs) are gaining prominence after bosonic ($I =
0$)~\cite{Cooper2018, Norcia2018b, Saskin2019, Covey2019, Norcia2019,
Madjarov2019, Wilson2019, Jackson2020, Madjarov2020, Young2020, Choi2021,
Burgers2021, Schine2021} and fermionic ($I > 0)$~\cite{Barnes2021, Jenkins2021,
Ma2021} isotopes recently joined this field. AEAs offer qualitative differences
and quantitative advantages over alkalis. For example, they offer long-lived
metastable states useful for applications including optical
metrology~\cite{Ludlow2015}; high-fidelity, lossless, state-resolved detection
via ``shelving''~\cite{Covey2019, Norcia2019, Madjarov2019}; and high-fidelity
Rydberg-mediated entanglement~\cite{Madjarov2020, Choi2021, Burgers2021}.

Fermionic isotopes have two potential advantages over their bosonic
counterparts: (1) their optical ``clock'' transition is significantly stronger
due to hyperfine mixing~\cite{Boyd2007}, and (2) the ground and metastable
``clock'' states have a nuclear spin degree of freedom decoupled from electronic
spin, which was recently utilized as a high-fidelity qubit~\cite{Barnes2021,
Ma2021, Jenkins2021}. These optical and nuclear degrees of freedom can be
identically trapped at a ``magic'' wavelength~\cite{Covey2019, Norcia2019,
Ye2008} where coherence times approach the minute scale~\cite{Young2020,
Barnes2021}. Such access to multiple highly coherent qubit types within a single
atom may obviate the need for heterogeneous qubit architectures, which have
become ubiquitous in myriad quantum science platforms~\cite{Schmidt2005,
Jiang2009, Pla2013, Arute2019, Singh2021}. We extend the term \textit{omg}
(``optical, metastable, and ground'') from a recent trapped ion
proposal~\cite{Allcock2021} to describe neutral fermionic AEAs in this context.

\begin{figure*}[t!]
    \centering
    \includegraphics[width=\textwidth]{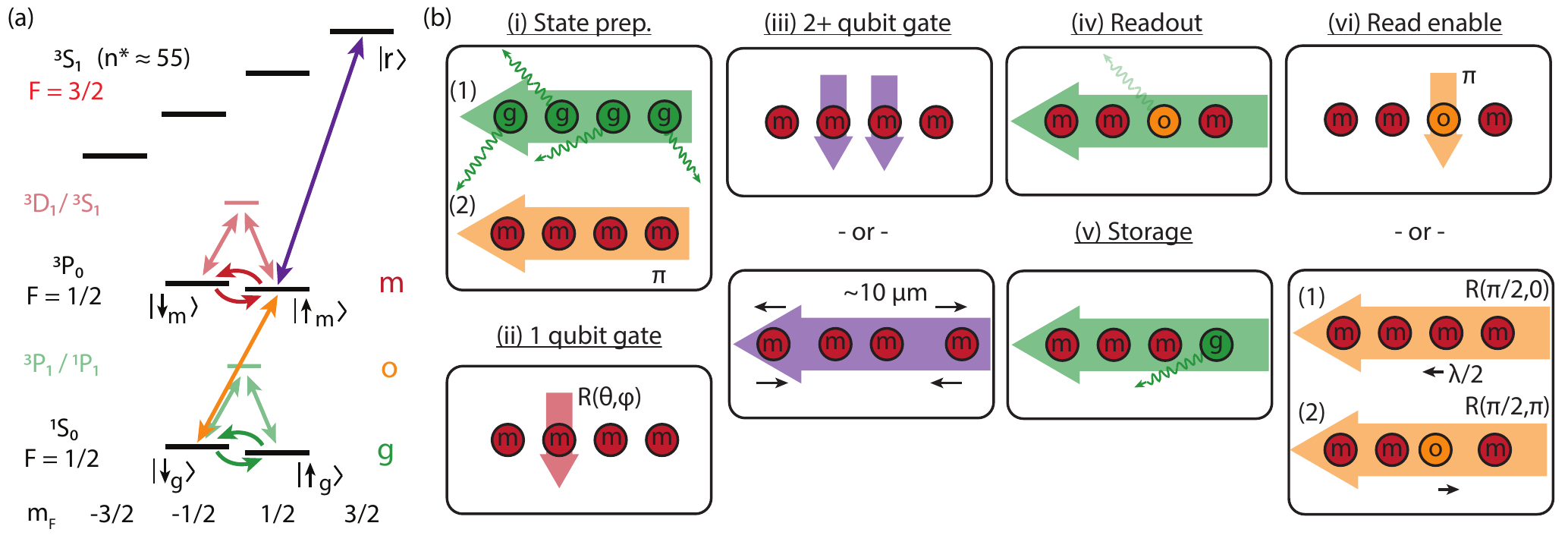}
    \caption{
        \textbf{Overview of the ${}^{171}\t{Yb}$ \textit{omg} architecture.} (a) The
        relevant level structure, showing the metastable (``m'') and ground (``g'') qubit
        encoded in the ``clock'' and ground states, respectively, as well as the ``clock'' (yellow arrow) and Rydberg (purple arrow) transitions. Qubit rotations
        can be performed with stimulated Raman processes via other, strong transitions (green and red arrows). (b) Overview of operations for ``m'' qubits. (i) State preparation begins with cooling and optical pumping (solid arrows are laser pulses and wiggly arrows are emitted photons) in the ground state via an auxiliary transition to ${}^1\t{P}_1$ or ${}^3\t{P}_1$ (green), followed by a global $\pi$-pulse on the optical qubit transition (yellow). (ii) Qubit rotations of ``m'' qubits (red) are performed with stimulated Raman transitions via other states. (The case is identical for ``g'' qubits). (iii) Rydberg-mediated two-qubit gates (purple), where targeted operations can be achieved in two ways (see text). (iv) Global readout is performed with the ``o'' qubit, where only the ``g'' sector fluoresces (translucent green) from light resonant with the auxiliary transition (green). (v) The ``m'' qubit can be used for storage that is immune to operations on the ``g'' qubit, including readout or rotations. (vi) The ``m'' qubit must be mapped to the ``o'' qubit (yellow) to perform readout with state-dependent fluorescence. There are two ways to perform a targeted read enable operation (see text).}
        \label{OMG}
\end{figure*}

\begin{figure}[t!]
    \centering
    \includegraphics[width=0.47\textwidth]{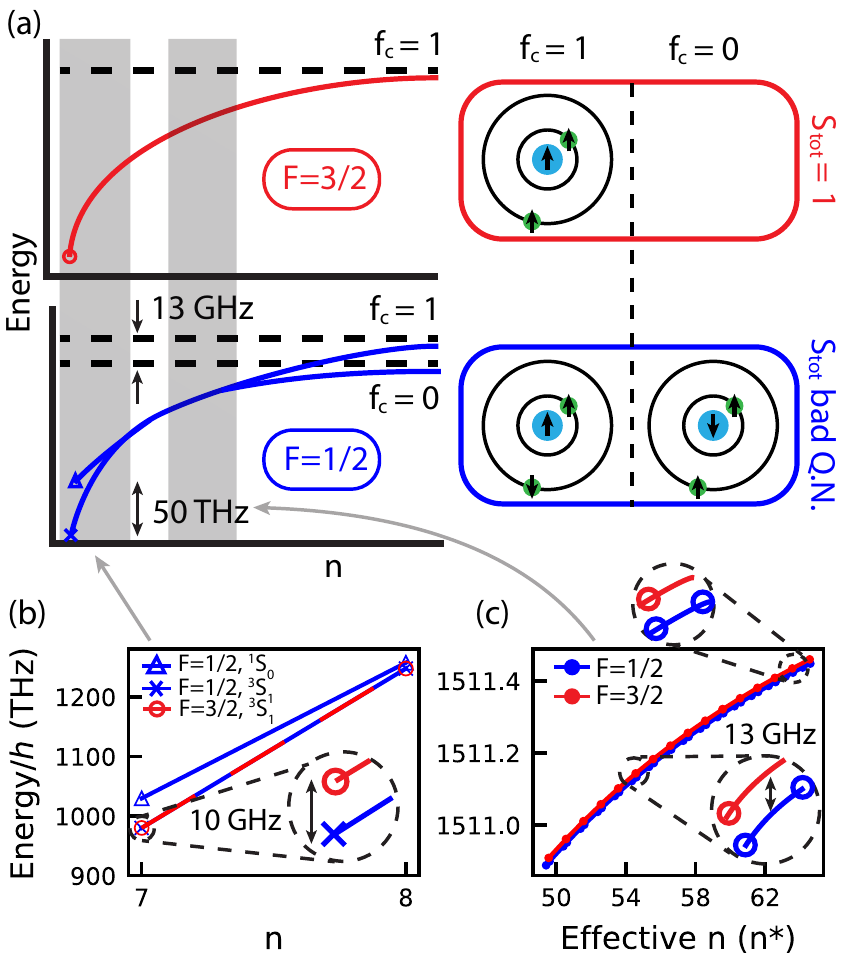}
    \caption{
        \textbf{The $S$-series Rydberg structure of ${}^{171}\t{Yb}$.} (a) The energy levels and spin configurations of the two
        $S$-series described by total angular momentum $F = \{1/2,\, 3/2\}$ versus principal quantum number ($n$). They
        asymptotically approach the hyperfine levels of the core ion $f_\t{c}
        = \{0,\, 1\}$ split by $12.6~\t{GHz}$. The $F = 3/2$ series is uniquely
        described by $S_\t{tot} = 1$ since all three spins must be aligned.
        The $F = 1/2$ series results from two configurations of the three spins,
        so $S_\t{tot}$ is not a good quantum number in this case. (b) The two series at
        small $n$, where the hyperfine splitting of the ${}^3\t{S}_1$ term into $F
        = \{1/2,\, 3/2\}$ is smaller than the singlet-triplet splitting. (c) The
        two series at $n^* \approx 50-65$ ($n \approx 55-70$) using multichannel
        quantum defect theory for the $F = 1/2$ series. The lower inset shows a separation of $\approx\Delta_\t{HFS}$ between the two series at $n^*\approx55$, while the upper inset shows a near degeneracy at $n^*\approx65$.
        \label{FigureRydbergHyperfine}
    }
\end{figure}

Here, we analyze the Rydberg-based \textit{omg} architecture for ${}^{171}\t{Yb}$ nuclear spins. We consider a ``g'' qubit $\{\downarrow_\t{g},\uparrow_\t{g}\}$ encoded in the ground state ${}^1\t{S}_0$ and a ``m'' qubit $\{\downarrow_\t{m}, \uparrow_\t{m}\}$ encoded in the metastable clock state ${}^3\t{P}_0$ [see Fig~\ref{FigureRydbergHyperfine}(a)]. These nuclear qubits can be manipulated by stimulated Raman transitions via other states~\cite{Barnes2021,Jenkins2021}, as is common for hyperfine qubits in neutral alkali atoms~\cite{Levine2019} and trapped ions~\cite{Allcock2021}. The ``g'' and ``m'' qubits are connected via the ``o'' qubit on the clock transition, and identical trapping conditions for all four states can be realized at the clock-magic wavelength of 759 nm where long coherence times are available~\cite{Ye2008,Ludlow2015}. We propose an architecture centered around the ``m'' qubit to leverage these degrees of freedom. We show that the combination of a modest magnetic field ($B \approx 200\,\t{G}$) and optical polarization intensity purity ($\approx 99\%$) is sufficient to perform $> 0.99$-fidelity operations on the nuclear qubits via the clock, Rydberg, and Raman transitions -- approaching the fault-tolerance threshold~\cite{Knill2005,Fowler2012}. We assume a coherence time of  $T_2^*\approx1$ sec, limited by mG-level magnetic field noise (see Appendix~\ref{MultilevelSim}) as well as off-resonant scattering from the tweezer traps (see Appendix~\ref{ClockShiftScatter}). This decoherence rate ($\sim2\pi\times1\,\t{Hz}$) can be compared to the limiting gate operation rate, the anticipated optical qubit Rabi frequency ($\Omega_\t{c} \approx 2\pi\times200\,\t{kHz}$), suggesting a promising platform for Rydberg-based entanglement in quantum computers and simulators~\cite{Daley2008, Gorshkov2009, Gorshkov2010, Pagano2019, Cong2021}, networks~\cite{Covey2019b, Huie2021}, and optical clocks~\cite{Gil2014, Kessler2014, Kaubruegger2019}.

\section{The \textit{omg} architecture}\label{sec: OMG}
There are countless ways to use the ``o'', ``m'', and ``g'' qubits, and the optimal variant of the \textit{omg} architecture depends critically on the application. For example, an optical atomic clock~\cite{Ludlow2015} with programmable entanglement~\cite{Schine2021} for achieving precision below the standard quantum limit~\cite{Gil2014,Kessler2014, Kaubruegger2019} will primarily focus on the ``o'' qubit. In this work, we focus on an architecture centered around the ``m'' qubit for three reasons: (1) the clock state is well suited for high-fidelity, single-photon coupling to Rydberg states in the ${}^3\t{S}_1$ series~\cite{Madjarov2020, Choi2021, Schine2021}, obviating the need for two-photon transitions limited by off-resonant scattering from the intermediate state~\cite{Levine2019,Ma2021}; (2) the clock state is well suited for shelving of quantum information during readout based on fluorescence from the ground state~\cite{Covey2019,Monz2016, Erhard2021}; and (3) the clock state of Yb has strong, telecom-band transitions to the $^3$D$_J$ series that offer opportunities for quantum networking~\cite{Covey2019b,Huie2021}.

For concreteness and to motivate the following analysis, we focus on the
operations shown in Fig.~\ref{OMG}(b). The ``m'' qubit will be used for
computation and storage. Qubits will be globally initialized by cooling and
optical pumping in the ground state via ${}^1\t{P}_1$ or ${}^3\t{P}_1$ followed
by a $\pi$-pulse on the clock transition to generate a fiducial register in
$\ket{\uparrow_\t{m}}$. Rotations of ``m'' qubits will be performed with
single-beam~\cite{Jenkins2021} stimulated Raman transitions via ${}^3\t{S}_1$ or
${}^3\t{D}_1$, which can be applied at the individual-qubit level via a
tightly-focused beam~\cite{Barnes2021} (see Appendix~\ref{AxialClockBeam} for an assessment of technical challenges with tightly-focused beams). Two- and multi-qubit gates will be
performed by coupling $\ket{\uparrow_\t{m}}$ to a Rydberg state $\ket{r}$, which
can be applied at the individual-qubit level with tightly-focused
beams~\cite{Graham2021} or with coherent transport techniques based on the
$1/r^6$ scaling of the Rydberg-Rydberg interactions that map proximity onto
connectivity~\cite{Lengwenus2010, Dordevic2021, Bluvstein2021}. Lossless,
state-resolved readout is performed by mapping qubits to the ``o''-type and then
collecting fluorescence from the ground state via its transition to
${}^1\t{P}_1$ or ${}^3\t{P}_1$~\cite{Covey2019, Norcia2019, Madjarov2020,
Barnes2021}. Fluorescence in the ``g'' manifold does not affect quantum
information in the ``m'' manifold, and thus storage in ``m'' enables parallel
processes in ``g'' such as single-qubit readout~\cite{Monz2016, Erhard2021} and
remote entanglement generation~\cite{Huie2021}.

Control of the ``o'' qubit plays a crucial role for type-casting and read-enabling. Single-qubit, mid-circuit readout requires a $\pi$-pulse on the clock transition to be performed at the individual qubit level. This can be accomplished with a tightly-focused beam or with a global~\cite{Madjarov2019,Norcia2019}, two-component pulse combined with coherent transport of the target qubit. Specifically, this technique would leverage the spatial variation of the optical phase combined with the ability to move a single atom by half a wavelength, corresponding to a $\pi$ phase shift, to perform a net $1\pi$-pulse on the target atom and a net $0\pi$-pulse on the spectator atoms~\cite{Chiaverini2004,Schaetz2004}. While we leave further analysis of this approach for future work, we note that the spatial precision available with adaptive optical elements such as Acousto-Optic Deflectors (AODs) is sufficient. Typical AOD-based tweezer systems have a position-to-radio frequency (RF) conversion of $\approx$10 $\mu$m/MHz~\cite{Endres2016,Covey2019}, and thus the $\approx$10 nm precision required for this protocol to be performed with a fidelity at the 0.99 fidelity level corresponds to only kHz-level precision of the RF signals.

\section{The Rydberg transition}\label{sec: Rydberg}
We now discuss the required operations of this architecture in detail, beginning with the Rydberg-based operations. Inspired by recent work~\cite{Madjarov2020, Choi2021, Schine2021}, we consider
Rydberg-mediated entanglement via the ${}^3\t{P}_0 \leftrightarrow {}^3\t{S}_1$
transition, where the latter has a principal quantum number of $n \approx 60$
[see Fig~\ref{OMG}(a)]. However, we note that a two-photon
transition from the ${}^1\t{S}_0$ ground state could be used
instead~\cite{Wilson2019, Burgers2021, Shi2021b, Ma2021} at the expense of
higher optical power and additional complexity, and was recently used to perform
two-qubit gates on the nuclear spin qubit in the ground state of
${}^{171}\t{Yb}$ at low field ($\approx 4\,\t{G}$)~\cite{Ma2021}. We require a protocol by which only one of the qubit states couples to the
Rydberg level~\cite{Graham2019, Levine2019}. Although we specifically consider
the ``m'' qubit, the requirements on the isolation of
the Rydberg drive from unwanted ``spectator'' states is stringent for all qubit
choices. The nuclear spins present a unique challenge due to their relatively
small energy splittings ($\approx \t{kHz/G}$). Hence, the development of a
high-fidelity two- or mutli-qubit gate protocol for fermionic AEAs will require
a detailed understanding of the Rydberg level structure~\cite{Robicheaux2018,
Ding2018, Robicheaux2019, Shi2021b, Ma2021}. We use multichannel quantum defect
theory~\cite{Aymar1996} (see Appendix~\ref{MQDT}) to gain new insight on this
structure. We consider $S$-series Rydberg states ($L = 0$), but our analysis can
be applied to $L > 0$.

The presence of a nuclear spin in an AEA creates a scenario that is
qualitatively different from both alkali and bosonic AEA Rydberg structures. In
the case of alkali species, the electron-nucleus coupling is small due to the
large orbit of the Rydberg electron, and thus the total electron angular
momentum $J$ is a good quantum number. In the case of bosonic AEAs, there are
two electron spins but no nuclear spin, so electron total spin $S$ (i.e. singlet
and triplet) and $J$ are good quantum numbers. Fermionic AEAs present a system
in which there are three coupled spins: two electrons and a nucleus. Indeed, the
hyperfine structure of the ionic core describes the Rydberg ionization
thresholds [see Fig.~\ref{FigureRydbergHyperfine}(a)]. The Rydberg series
corresponding to total angular momentum $F = 1/2$ is not well described by
$S_\t{tot}$ -- meaning that the singlet/triplet designation is inappropriate
-- since two configurations ($f_\t{c} = 0$ and $f_\t{c} = 1$) both
contribute, and a multichannel quantum defect theory~\cite{Aymar1996} is
required. Conversely, the series corresponding to $F = 3/2$ can only be obtained
from one configuration ($f_\t{c} = 1$) and is thus well described by
$S_\t{tot} = 1$. Due to its clean structure for all $n$ (assuming no
perturbers) and its designation as a ``spin triplet,'' we target this $F = 3/2$
series as being ideally suited for our two- or multi-qubit entangling
operations~\cite{Ma2021}.

\begin{figure}[t!]
    \centering
    \includegraphics[width=0.47\textwidth]{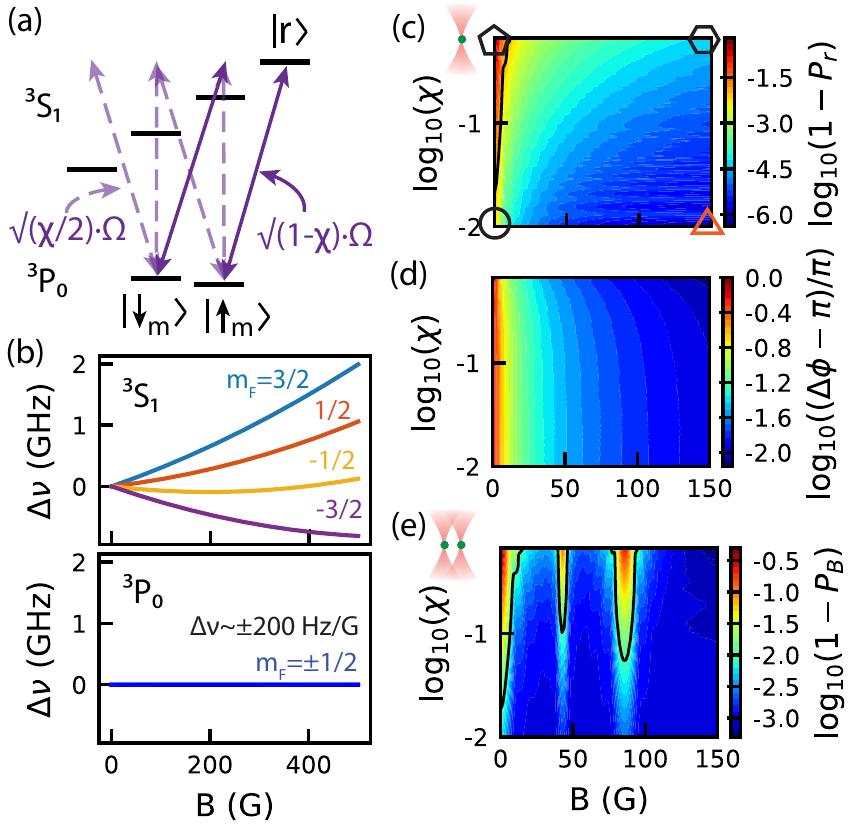}
    \caption{
        \textbf{Analysis of the Rydberg transition.} (a) The six-level system
        showing the nuclear qubit $\{\downarrow_\t{m},\, \uparrow_\t{m}\}$ in
        the clock state and the four $m_F$ states in the $F = 3/2$ Rydberg
        state. We target the $\sigma^+$ ``stretched'' transition
        $\ket{\uparrow_\t{m}} \leftrightarrow \ket{r}$, but imperfect
        polarization creates off-resonant couplings to other states. We
        parameterize the strengths of these couplings with $\sqrt{\chi/2}$,
        since polarization intensity purity is associated with optical power,
        $P$, and $\Omega \sim \sqrt{P}$. Weighting by Clebsch-Gordan
        coefficients is included. (b) The magnetic field maps of the clock
        (including hyperfine interaction~\cite{Boyd2007}) and Rydberg (including
        the diamagnetic shift~\cite{Madjarov2020}) states. (c) Single-atom
        $\pi$-pulse infidelity, initialized in $\ket{\uparrow_\t{m}}$, under
        various polarization impurities $\chi$ and magnetic fields $B$. The
        color scale is the population not in $\ket{r}$, $1-P_r$. The shapes
        indicate the conditions under which Rabi oscillations are shown in
        Fig.~\ref{FigureRydbergRabi}. (d) Single-atom relative phase accrual on
        the $\{\downarrow_\t{m},\, \uparrow_\t{m}\}$ qubit resulting from a
        $2\pi$-pulse on the $\ket{\uparrow_\t{m}} \leftrightarrow \ket{r}$
        transition under various $\chi$ and $B$. The color scale shows the phase
        accrual in units of $\pi$ radians, where $\pi$ is expected in the ideal
        case. (e) Two-atom $\pi$-pulse infidelity, initialized in
        $\ket{\uparrow_\t{m} \uparrow_\t{m}}$, under various $\chi$ and $B$.
        The color scale is the population not in the $\ket{B}$ Bell state,
        $P_B$ (see text). The black lines in (c) and (e) show where $P = 0.99$.
        \label{FigureRydbergDrive}
    }
\end{figure}

Figure~\ref{FigureRydbergHyperfine}(b) and (c) shows the spectrum of the $F =
1/2$ and $F = 3/2$ series of the $S$ manifold at low principal quantum number
$n$ and effective principal quantum number near $n^* \approx 55$, respectively. In
the small-$n$ limit~\cite{Berends1992}, the singlet-triplet splitting is much
larger than the hyperfine splitting of $F = \{1/2,\, 3/2\}$ in the ${}^3\t{S}_1$
manifold ($\approx 10\,\t{GHz}$~\cite{Berends1992}). Near $n^*=55$, the
two configurations of $F = 1/2$ -- analyzed with multichannel quantum defect
theory~\cite{Aymar1996} (see Appendix~\ref{MQDT}) -- follow the same trend line
before separating to asymptotically approach the $f_\t{c} = \{0,\, 1\}$ limits
[see Fig.~\ref{FigureRydbergHyperfine}(a)]. The $F = 3/2$ series has only a
single configuration asymptotically approaching $f_\t{c} = 1$. The state
energies in this series can thus be modeled using the known energies of the
${}^3\t{S}_1$ series in the bosonic isotope ${}^{174}\t{Yb}$ (obtained from
Ref.~\cite{Wilson2019}) plus the hyperfine splitting $\Delta_\t{HFS} = 2\pi\times
12.6\,\t{GHz}$ of the ${}^{171}\t{Yb}$ ionic core (see Appendix~\ref{3S1}).
Figure~\ref{FigureRydbergHyperfine}(c) shows both the $F = 1/2$ and $F = 3/2$
series near $n^* = 55$, where the figure of merit is the energy separation
between the two series and the associated resolvability of a given state. Near $n^*=55$ (lower inset), the
$\approx 13\,\t{GHz}$ separation of the states in the $F = 3/2$ series
from the closest ones in the $F = 1/2$ series suggests excellent isolation in
the presence of strong laser coupling. However, there are near-degeneracies between the two series, such as near $n^*=65$ (upper inset), that must be avoided. This is quantified more precisely by Lu-Fano plots~\cite{Lu1970} of the two series (see Appendix~\ref{MQDT}).

\begin{figure}[t!]
    \centering
    \includegraphics[width=0.47\textwidth]{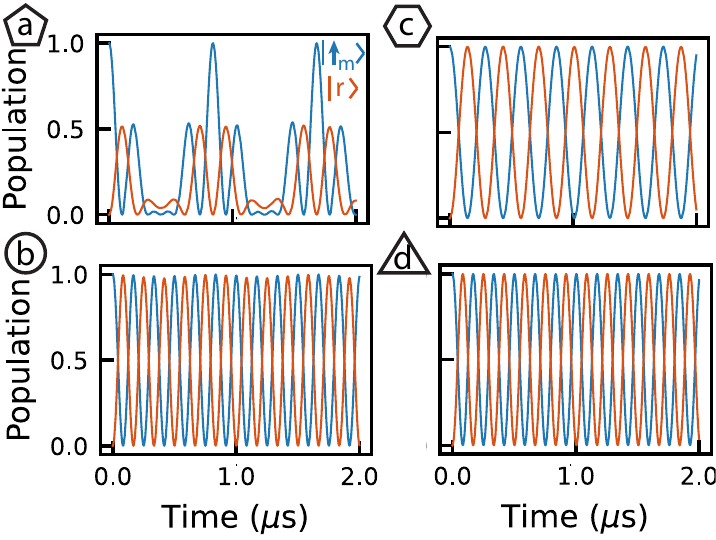}
    \caption{
        \textbf{Single-atom Rydberg Rabi oscillations.} Populations
        $\ket{\uparrow_\t{m}}$ (blue) and $\ket{r}$ (orange) versus time under
        various $\chi$ and $B$ conditions. Note that population is not conserved
        due to leakage to other states in the six-state system when $\chi>0$.
        (a) $\chi=2/3$ (fully unpolarized) and $B = 0\,\t{G}$. (b) $\chi =
        10^{-2}$ and $B = 0\,\t{G}$. (c) $\chi=2/3$ and $B = 150\,\t{G}$. (d)
        $\chi = 10^{-2}$ and $B = 150\,\t{G}$. Note that dephasing mechanisms
        (see text) are not included to avoid obfuscating the atomic structure
        considerations.
        \label{FigureRydbergRabi}
    }
\end{figure}

We consider the use of the $\sigma^+$-polarized ``stretched'' transition between
${}^3\t{P}_0$ $\ket{m_F = 1/2} \equiv \ket{\uparrow_\t{m}}$ and ${}^3\t{S}_1$
($n \approx 60$) $\ket{m_F = 3/2} \equiv \ket{r}$ [see
Fig.~\ref{FigureRydbergDrive}(a)] to obviate the coupling with
$\ket{\downarrow_\t{m}}$ in the presence of a slight polarization impurity
[dashed arrows in Fig.~\ref{FigureRydbergDrive}(a)]. (See Appendix~\ref{Linear}
for analysis of the $\pi$-polarized case.) The nature of the $F = 3/2$ series
allows for the standard Land\'e g-factors to be used to compute Zeeman
splittings. We find $\Delta_\t{Z}/(2\pi) = m_F \times 1.9\,\t{MHz/G}$ in the low-field
limit and we include the well-known~\cite{Madjarov2020} $m_F$-independent
diamagnetic interaction $\Delta_\t{DM} \sim |\mathbf{d} \times \mathbf{B}|^2$ that
dominates at $B \gtrsim 800\,\t{G}$. We neglect
hyperfine mixing between Rydberg manifolds as there is no significant contribution for the conditions considered here (see Appendix~\ref{3S1}).
The magnetic field shifts of the Rydberg states and the ${}^3\t{P}_0$ clock
states are shown in Fig.~\ref{FigureRydbergDrive}(b).

To assess the prospect of gate operations on the $\ket{\uparrow_\t{m}}
\leftrightarrow \ket{r}$ transition, we numerically simulate a drive of
strength $\Omega_\t{R} = 2 \pi \times 6\,\t{MHz}$ on the six-level system (see
Appendix~\ref{MultilevelSim}) for various magnetic fields $B$ and polarization
intensity impurities $\chi$ [defined in Fig.~\ref{FigureRydbergDrive}(a)]. We assume magnetic field magnitude and orientation uniformity at the $10^{-4}$ level in a well-designed Helmholtz field~\cite{Moses2015,Covey2016}. This value of $\Omega_\t{R}$ was chosen based on similar work with global Rydberg pulses for strontium~\cite{Madjarov2020}; tightly-focused pulses would require less power, albeit with added technical challenges (see Appendix~\ref{AxialClockBeam}). The $\pi$-pulse infidelity (population not in $\ket{r}$, $1-P_r$) for a single atom is shown
in Fig.~\ref{FigureRydbergDrive}(c), where even 90\% polarization intensity
purity ($\chi = 10^{-1}$) at $B = 100\,\t{G}$ gives a transfer fidelity of
$\mathcal{F} \approx 0.99$. The shapes included in
Fig.~\ref{FigureRydbergDrive}(c) denote the plots in
Fig.~\ref{FigureRydbergRabi} showing Rabi oscillations during a prolonged pulse
under those conditions.

We also consider the accrued relative phase on the $\{\downarrow_\t{m},\,
\uparrow_\t{m}\}$ qubit due to the undesired couplings during a
$\ket{\uparrow_\t{m}} \leftrightarrow \ket{r}$ pulse. Although finite phase
accrual due to light shifts during gates can be tolerated, fluctuations in this
phase due to, e.g., intensity fluctuations can have deleterious effects on the
quantum circuit. To obviate this problem, it is clearly optimal to minimize the
phase accrual due to parasitic couplings. To probe this effect in our system, we
consider the accrued phase during a $2 \pi$ pulse on $\ket{\uparrow_\t{m}}
\leftrightarrow \ket{r}$ (see Appendix~\ref{PhaseAccrual}) for various magnetic
fields and polarization intensity impurities [Fig.~\ref{FigureRydbergDrive}(d)].
We find that the accrued phase relative to the ideal case with zero coupling to
other Rydberg states, is $\Delta \phi \lesssim 0.01 \pi$ for $B \gtrsim
150\,\t{G}$ for a wide range of $\chi$. Percent-level fluctuations in this phase
are negligible, and the phase itself is expected to be sufficiently small for a
fidelity approaching 0.99 in intolerant applications.

Finally, we consider the prospects for two-qubit entanglement. Although we are
interested in entanglement of low-lying states such as $\{\downarrow_\t{m},\,
\uparrow_\t{m}\}$ via protocols such as in Refs.~\cite{Levine2019, Graham2019,
Schine2021, Martin2021}, we consider only the $\{\uparrow_\t{m},\, r\}$ qubit
here since operation of this transition is required in any protocol and thus
presents a fidelity limit. We look at the pulse fidelity in the two-atom case,
assuming $C_6(n^* = 55) \approx 300\,\t{GHz} \cdot \t{$\mu$m}^6$ based on
recently measured values~\cite{Ma2021} that give a Rydberg interaction shift
$U_\t{VdW} / \hbar \approx 2 \pi \times 160\,\t{MHz}$ ($\approx 27 \Omega$) for
an inter-atom separation of $r = 3.5\,\t{$\mu$m}$ -- deep within the Rydberg
blockade limit. We consider the entangled ``bright'' Bell state $\ket{B} \equiv
(\ket{\uparrow_\t{m} r} + \ket{r \uparrow_\t{m}}) / \sqrt{2}$, where the two
elements in the state refer to the two atoms~\cite{Levine2018, Madjarov2020}. We
study the population not in $\ket{B}$, $1-P_B$, after a $\pi$-pulse from
$\ket{\uparrow_\t{m}\, \uparrow_\t{m}}$ to $\ket{B}$ for various magnetic fields
and polarization intensity purities [Fig.~\ref{FigureRydbergDrive}(e)].
Resonances with the Rydberg states $\ket{\{r_\Downarrow,\, r_\downarrow,\,
r_\uparrow\}} \equiv \ket{m_F = \{-3/2,\, -1/2,\, +1/2\}}$ occur at magnetic
fields where $U_\t{VdW} = [\Delta_\t{z}(m_F = 3/2) - \Delta_\t{z}(m_F)]\times
B$. The resonances corresponding to $\ket{r_\downarrow}$ and $\ket{r_\uparrow}$
manifest in Fig.~\ref{FigureRydbergRabi}(e) as regions with low pulse fidelity,
exacerbated by high $\chi$, while the resonance with $\ket{r_\Downarrow}$ is not
apparent only because the initial state $\ket{\uparrow_\t{m} \uparrow_\t{m}}$
does not couple to it. This effect is irrelevant at fields of $B \gtrsim
200\,\t{G}$ that we later identify as optimal, and can be entirely removed by
instead driving the $\ket{\downarrow_\t{m}} \leftrightarrow \ket{r_\Downarrow}$
($\sigma^+$) transition since $U_\t{VdW} > 0$.

This analysis suggests that our nuclear spin qubit is a viable platform for
quantum science with Rydberg states, enabling two-qubit entanglement and
many-body dynamics at or beyond the current fidelity record~\cite{Bernien2017,
Levine2019, Madjarov2020, Choi2021}. We briefly consider in Appendix~\ref{RydbergNoise} the
well-known limitations to coherent Rydberg excitation: laser frequency noise, finite Rydberg state lifetime, DC Stark and Zeeman shifts, and random Doppler shifts due to finite atom temperature. There are also challenges associated with
individual-qubit addressing~\cite{Barnes2021, Graham2021}, which we consider in Appendix~\ref{AxialClockBeam}. However, these technical
limitations are ubiquitous across species and qubit encodings and are thoroughly addressed elsewhere~\cite{Levine2018,
deLeseleuc2018, Wilson2019,Madjarov2020}, but some are perhaps easier to mitigate with
AEAs due to their access to higher Rydberg-excitation Rabi frequencies and
colder temperatures~\cite{Madjarov2020, Burgers2021, Schine2021, Jenkins2021}.
The point of this analysis is rather to demonstrate that the nuclear spin qubit
is not limited by atomic structure under the correct conditions.

\section{The clock transition}\label{sec: Clock}
\begin{figure}[t!]
    \centering
    \includegraphics[width=0.5\textwidth]{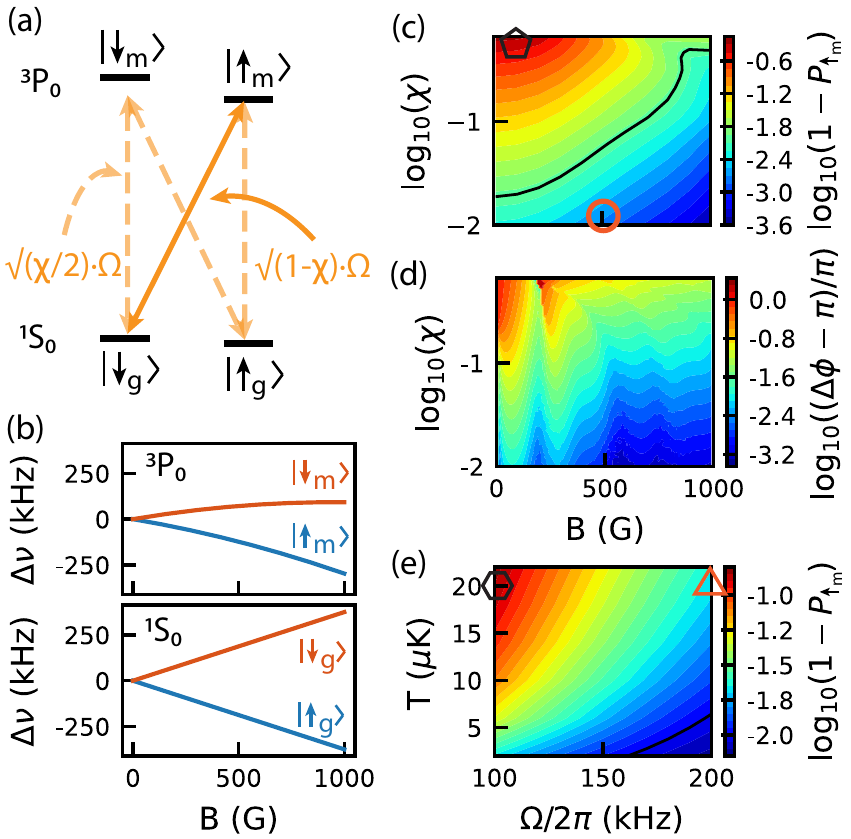}
    \caption{
        \textbf{Analysis of the clock transition.} (a) The four-level system
        showing the nuclear qubits $\{\downarrow_\t{g},\, \uparrow_\t{g},\,
        \downarrow_\t{m},\, \uparrow_\t{m}\}$ in the ground and clock states. We
        target the $\sigma^+$ ``stretched'' transition $\ket{\downarrow_\t{g}}
        \leftrightarrow \ket{\uparrow_\t{m}}$, but imperfect polarization
        creates off-resonant couplings to other states. We parameterize this
        identically to the Rydberg case in Figure~\ref{FigureRydbergDrive}. (b)
        The magnetic field maps of the ground and clock states (including
        hyperfine interaction~\cite{Boyd2007}). (c) $\pi$-pulse infidelity,
        initialized in $\ket{\downarrow_\t{g}}$, under various $\chi$ and $B$.
        The shapes indicate the conditions under which Rabi oscillations are
        shown in Fig.~\ref{FigureClockRabi}. The color scale is the population
        not in $\ket{\uparrow_\t{m}}$, $1-P_{\uparrow_\t{m}}$, and the black
        line shows where $P_{\uparrow_\t{m}} = 0.99$. (d) Relative phase accrual
        on the $\{\downarrow_\t{g},\, \uparrow_\t{g}\}$ qubit resulting from a 
        $2\pi$-pulse on the $\ket{\uparrow_\t{g}} \leftrightarrow
        \ket{\downarrow_\t{m}}$ transition under various $\chi$ and $B$. The
        color scale shows the phase accrual in units of $\pi$ radians, where
        $\pi$ is expected in the ideal case. (e) Infidelity due to finite
        temperature effects in the four-level system (not including phase
        noise). $1 - P_{\uparrow_\t{m}}$ versus temperature and Rabi frequency.
        The shapes refer to Fig.~\ref{FigureClockRabi}.
        \label{FigureClockDrive}
    }
\end{figure}

We now turn to a discussion of the optical clock transition. As
discussed above, global clock pulses are needed for initialization in the ``m''
qubit. Also, targeted clock operations for read-enabling can be performed with
either tightly-focused clock beams, or potentially with global pulses combined with targeted position shifts~\cite{Chiaverini2004,Schaetz2004}. For the
purposes of this discussion, the most important parameters are the Rabi
frequency $\Omega_\t{c}$ and the trapping frequency $\omega$ along the $k$-vector
of the clock pulse. In either case we assume $\Omega_\t{c} \approx 2 \pi \times
200\,\t{kHz}$ is realistic (see Appendix~\ref{ClockShiftScatter}), which naturally requires more optical power in the
global addressing case. Specifically, based on the well-known transition
strength~\cite{Hong2005}, $P = 50\,\t{mW}$ would be required for a beam of waist
radius $w_0 = 20\,\mu\t{m}$ aligned along a one-dimensional
array~\cite{Norcia2019, Madjarov2019}. The relevant level structure is shown in
Fig.~\ref{FigureClockDrive}(a). We again choose to drive a $\sigma^+$-transition
to limit the possible undesired couplings. The Zeeman energies of the nuclear
states are shown in Fig.~\ref{FigureClockDrive}(b), where hyperfine interactions
affect the trend in the ${}^3\t{P}_0$ state. The differential g-factor at low
field is $\approx 200 \,\t{Hz/G}$ (see Appendix~\ref{HFMix}), so we are reliant
on polarization selectivity since the drive bandwidth will exceed the energy
separation.

We analyze a $\pi$-pulse of the clock transition, initialized in
$\ket{\downarrow_\t{g}}$, for various polarization intensity purities and
magnetic fields. In Appendix~\ref{PhaseNoise}, we consider phase noise since it
constitutes a liability unique to optical qubits. However, we neglect phase
noise here to avoid obfuscating the internal dynamics and to keep the results
general. Fig.~\ref{FigureClockDrive}(c) shows the population not in
$\ket{\uparrow_\t{m}}$, $1 - P_{\uparrow_\t{m}}$, and we find that a field
strength of $B \gtrsim 200\,\t{G}$ with $\chi \gtrsim 10^{-2}$ polarization
intensity purity is sufficient for population transfer exceeding 0.99. As a more
stringent requirement than the $\pi$-pulse fidelity, we again consider relative
phase accrual, now on the $\{\downarrow_\t{g},\, \uparrow_\t{g}\}$ qubit,
resulting from undesired couplings (see Appendix~\ref{PhaseAccrual}).
Specifically, we consider a $2 \pi$-pulse on the $\ket{\downarrow_\t{g}}
\leftrightarrow \ket{\uparrow_\t{m}}$ transition. We find a relative phase
accrual of $\Delta \phi \lesssim 0.01 \pi$ for $B \gtrsim 200$ G and $\chi
\gtrsim 10^{-2}$, sufficient for operations with a fidelity of $\gtrsim 0.99$.

Finally, the analysis in Figs.~\ref{FigureClockDrive}(c) and (d) was performed
without considering motional degrees of freedom. We now consider finite
temperature and atomic motion effects. For concreteness, we now
assume a global pulse with k-vector along the radial direction of the tweezer
traps. In Appendix~\ref{AxialClockBeam}, we consider pulses propagating in the
axial direction, where the performance in this respect is improved due to the
larger disparity between $\Omega_\t{c}$ and $\omega$. Other technical challenges naturally
emerge, however (see Appendix~\ref{AxialClockBeam}). We assume a radial trap
frequency in the tweezer of $\omega_r = 2 \pi \times 70\,\t{kHz}$ (corresponding
to a tweezer with $1/e^2$ waist radius of 700 nm and depth of 500 $\mu$K), which
is significantly smaller than the Rabi frequency $\Omega_\t{c} = 2 \pi \times
200\,\t{kHz}$. These trap conditions correspond to a Lamb-Dicke parameter of
$\eta_r = 0.22$, where $\eta = k x_0$ depends on the wavenumber $k$ of the
driving laser and the harmonic oscillator length $x_0 = \sqrt{\hbar / (2
m_\t{Yb} \omega_r)}$ of the atom in the trap. In the $\Omega_\t{c} \gg \omega_r$
limit with ``magic'' trapping conditions (under which the trap frequency in the
ground and clock state are equal~\cite{Ye2008, Covey2019}), we choose the basis
states~\cite{Kale2020, Jenkins2021} to be $\ket{g, n} = \ket{g} \otimes \ket{n}$
and $\ket{e, \xi(n)} = \ket{e} \otimes e^{i \eta(\hat{a} + \hat{a}^\dagger)}
\ket{n}$, where $g$ ($e$) are the electronic ground (excited) state and $n$ is
the motional quantum number. We perform this analysis with all four states in
the ground-clock manifold, but only list two here for brevity. This basis
greatly simplifies the calculation for the case of a strong driving field since
the Hamiltonian becomes sparse. See Appendix~\ref{Thermal} for details.

\begin{figure}[t!]
    \centering
    \includegraphics[width=0.5\textwidth]{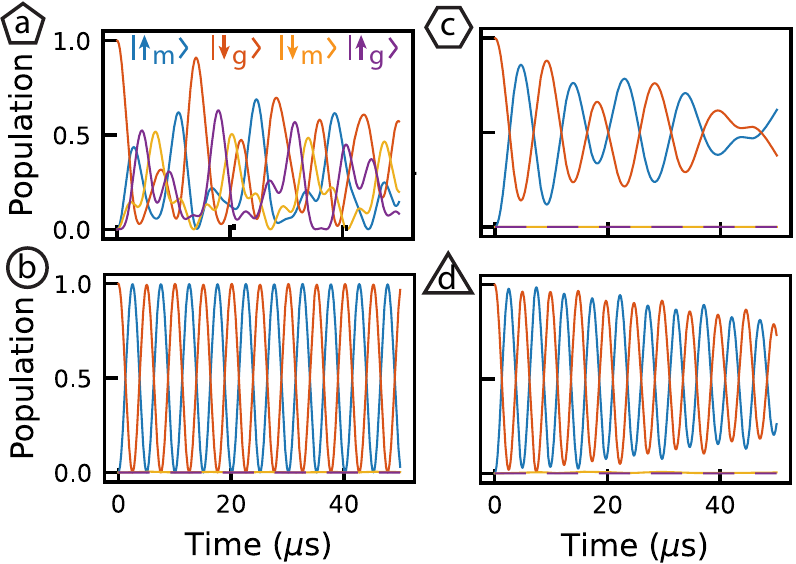}
    \caption{
        \textbf{Clock Rabi oscillations.} The population in
        $\ket{\uparrow_\t{m}}$ (blue) and $\ket{\downarrow_\t{g}}$ (orange)
        versus time under various $\chi$ and $B$ conditions. Note that
        population in these two states is not conserved as it leaks to
        $\ket{\uparrow_\t{g}}$ (purple) and $\ket{\downarrow_\t{m}}$ (yellow)
        when $\chi > 0$. (a) $\chi = 2/3$, $B = 100\,\t{G}$, and $\Omega / 2 \pi
        = 200\,\t{kHz}$; no motion. (b) $\chi = 10^{-2}$, $B = 500\,\t{G}$, and
        $\Omega / 2 \pi = 200\,\t{kHz}$; no motion. (c) $\chi = 10^{-2}$, $B =
        500\,\t{G}$, and $\Omega / 2 \pi = 100\,\t{kHz}$; with motion ($T =
        20\,\mu\t{K}$). (d) $\chi = 10^{-2}$, $B = 500\,\t{G}$, and $\Omega / 2
        \pi = 200\,\t{kHz}$; with motion ($T = 20\,\mu\t{K}$).
        \label{FigureClockRabi}
    }
\end{figure}

At $B = 500\,\t{G}$ and $\chi = 10^{-2}$, we study the dependence of the
$\pi$-pulse fidelity on temperature over the range of $T \in [2, 20]\,\mu\t{K}$
(where temperatures of $T \lesssim 5\,\mu\t{K}$ are expected~\cite{Covey2019,
Norcia2019, Jenkins2021, Ma2021}), studied for Rabi frequencies $\Omega \in 2
\pi \times [100, 200]\,\t{kHz}$ [see Fig.~\ref{FigureClockDrive}(e)].
Intuitively, higher $\Omega_\t{c}$ is more forgiving of higher $T$, and we
predict pulse fidelities exceeding 0.99 with $\Omega_\t{c} = 2 \pi \times
200\,\t{kHz}$ for $T \lesssim 10\,\t{$\mu$K}$. Note that although we focus here
on a single, relatively high trap frequency~\cite{Norcia2019, Madjarov2019,
Young2020, Barnes2021}, the situation improves with lower $\omega_r$, as shown
nicely in Ref.~\cite{Jenkins2021}. Conceptually, a lower trap frequency gives
slower atomic motion which decreases the Doppler shift.
Figure~\ref{FigureClockRabi} shows Rabi oscillations under the conditions
indicated with shapes in Fig.~\ref{FigureClockDrive}. Indeed, we find the limit
$\Omega_\t{c} \gg \omega_r$ to be relatively immune to thermal effects, as shown
for $T = 20\,\mu\t{K}$ in Fig.~\ref{FigureClockRabi}(d). Note that, depending on
$B$, the $\pi$-pulse fidelity will begin to decrease with increasing $\Omega$
simply because of the increasing coupling to the ``spectator'' states. We study
this interplay of $\Omega$ and $B$ in Appendix~\ref{ClockRabi}.

\section{Single-beam Raman gates}\label{sec: Raman}
\begin{figure}[t!]
    \centering
    \includegraphics[width=0.47\textwidth]{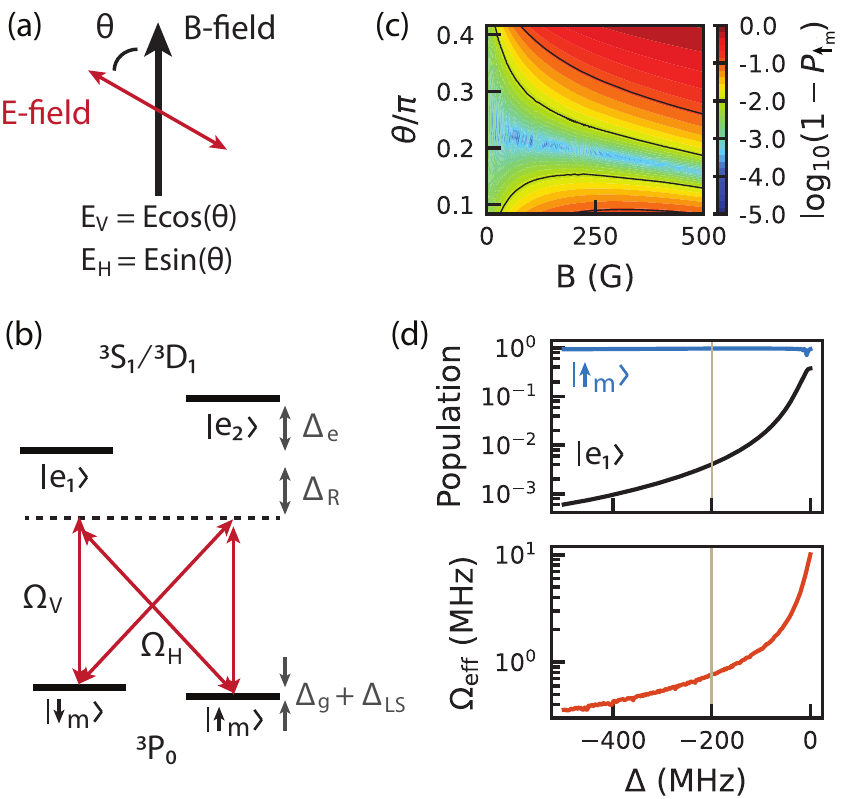}
    \caption{
        \textbf{Single-beam Raman gates.} (a) The polarization axis (electric
        field) of the linearly-polarized Raman beam with respect to the
        quantization axis (magnetic field), given by $\theta$, determines the
        projections of the electric field onto the vertical and horizontal axes
        (parallel and perpendicular to the magnetic field, respectively). (b)
        The ``m'' qubit in the clock state is off-resonantly coupled to an
        additional excited state (${}^3\t{S}_1$ or ${}^3\t{D}_1$) via its $F = 1/2$
        level. The vertical and horizontal components of the electric field
        drive $\pi$ and $\sigma$ transitions with Rabi frequencies
        $\Omega_\t{V}$ and $\Omega_\t{H}$, respectively (see
        Appendix~\ref{MultilevelSim}). (c) The $\pi$-pulse infidelity of the
        ``m'' qubit versus $\theta$ and magnetic field with constant drive
        strength of the Raman beam, detuned from $\ket{e_1}$ by $\Delta_R = 2
        \pi \times 200\,\t{MHz}$. The ``magic'' region showing improved transfer
        depends on $B$ and narrows for larger fields. The contour lines show the
        $10^{-2}$ and $10^{-3}$ infidelity levels. (d)
        $\Omega_\t{eff}$ and $P_{\uparrow_\t{m}}$ and $P_{e_1}$ (populations in
        the target ``m'' state and intermediate state $\ket{e_1}$, respectively)
        after a $\pi$-pulse versus $\Delta_R$ for constant intensity
        corresponding to $\Omega_\t{V} = 2 \pi \times 20\,\t{MHz}$ when $\theta
        = 0$.
        \label{FigureRaman}
    }
\end{figure}

We now turn to a discussion of rotations of the nuclear qubits
(single-qubit gates) via stimulated Raman pulses. Inspired by recent work
demonstrating Raman-based control of the ``g'' qubit~\cite{Jenkins2021}, we
focus on single-beam Raman gates. Crucially, the splitting of the nuclear qubits
is much smaller -- even in modest field ($\lesssim \t{kHz/G}$) -- than the
target effective Rabi frequency of $\Omega_\t{eff} \approx 2 \pi \times
1\,\t{MHz}$~\cite{Levine2019, Allcock2021, Jenkins2021}. Thus, a single beam
with a linear polarization tilted by an angle $\theta$ with respect to the
quantization axis (magnetic field) [see Fig.~\ref{FigureRaman}(a)] can provide
components that drive both the $\pi$- and $\sigma$-transitions of the Raman
coupling~\cite{Jenkins2021} [see Fig.~\ref{FigureRaman}(b)]. As shown in
Fig.~\ref{OMG}(a), the ``g'' and ``m'' qubits can be controlled identically,
only via a different intermediate state. For clarity, and to match our proposed
architecture shown in Fig.~\ref{OMG}(b), we focus on the ``m'' qubit which can
be controlled via the ${}^3\t{D}_1$ or ${}^3\t{S}_1$ state.

The analysis of the clock transition suggests that operation at
a magnetic field of $\gtrsim 200\,\t{G}$ is required for effective
implementation of the \textit{omg} architecture. This large field significantly
affects the gate operation -- not because of the Zeeman shift of ``m'' qubit
itself, but because of the shift on the intermediate state which has electronic
angular momentum ($\sim \t{MHz/G}$). Crucially, the detuning from the
intermediate state $\Delta_R$ is approximately equal to the Zeeman splitting on
its sublevels $\Delta_e$ under the conditions considered here [see
Fig.~\ref{FigureRaman}(b)]. Therefore, the use of a $F = 3/2$ level for the
intermediate state at high field would be drastically different than the
low-field case studied recently where $\Delta_R \gg
\Delta_e$~\cite{Jenkins2021}.

Instead, we consider a $F = 1/2$ level, which is available for
both ${}^3\t{S}_1$ and ${}^3\t{D}_1$. At a high field where $\Delta_R \approx
\Delta_e$, the intermediate state coupling via $m_F = -1/2$ is significantly
stronger than via $m_F = 1/2$ for a red-detuned laser, which also means that
$m_F = -1/2$ contributes more to the light shifts on the ``m'' qubit from the
Raman pulse. This presents a unique opportunity: we predict a ``magic''
polarization angle $\theta = \theta_m(B)$ for which the differential light shift
on the ``m'' qubit, $\Delta_\t{LS}$, exactly cancels its Zeeman splitting,
$\Delta_g$. Performing the Raman gates at this magic angle constitutes a pure
$\sigma_x$ rotation on the Bloch sphere, obviating complications due to the
inevitable $\alpha \sigma_x + \beta \sigma_z$ nature of the rotation when
$\Delta_g$ is left uncompensated. Fig.~\ref{FigureRaman}(c) shows the
$\pi$-pulse infidelity for the ``m'' qubit  versus $\theta$ and the magnetic
field, clearly showing excellent transfer in a region around $\theta_m(B)$ that
narrows as $B$ increases. This data uses $\Delta_R = 2 \pi \times 200\,\t{kHz}$,
and $\Omega_\t{eff} \approx 2 \pi \times 1\,\t{MHz}$.
Figure~\ref{FigureRaman}(d) shows $\Omega_\t{eff}$ and $P_{\uparrow_\t{m}}$ and
$P_{e_1}$ (populations in the target ``m'' state and intermediate state
$\ket{e_1}$, respectively) after a $\pi$-pulse versus $\Delta_R$ for constant
intensity corresponding to $\Omega_\t{V} = 2 \pi \times 20\,\t{MHz}$ when
$\theta = 0$ (See Appendix~\ref{MultilevelSim}). Technical challenges for targeted gates with a tightly-focused beam are considered in Appendix~\ref{AxialClockBeam}.

We focus on $\Delta_R = 2 \pi \times 200\,\t{kHz}$, which is
used in Fig.~\ref{FigureRaman}(c). As shown in Fig.~\ref{FigureRaman}(d), the population in $|$$e_1\rangle$ is $P_{e_1}\approx 4\times10^{-3}$ under this value of $\Delta_R$. Since the intermediate state has a total decay rate
of $\Gamma_e \approx 2 \pi \times 500\,\t{kHz}$ for ${}^3\t{D}_1$, the effective
scattering rate from $|$$e_1\rangle$ is $\Gamma_e^\t{eff} \lesssim 2 \pi \times
2\,\t{kHz}$, which we should compare to $\Omega_\t{eff} \lesssim 2 \pi \times
1\,\t{MHz}$, suggesting that Raman $\pi$-pulse fidelities well above 0.99 are
possible. Arbitrary rotations on the Bloch sphere can be accomplished by using
additional pulses with $\theta = 0$ or $\pi / 2$ such that there is no Raman
condition and the pulse only provides a light shift for the
$\ket{\downarrow_\t{m}}$ or $\ket{\uparrow_\t{m}}$ state, respectively, thereby
providing a controlled $\sigma_z$ rotation~\cite{Jenkins2021}.

\section{Conclusion and outlook}\label{sec: Conclusion}
This analysis demonstrates that the structure of ${}^{171}\t{Yb}$ is well suited
for high-fidelity quantum circuits featuring multiple qubit modalities within
the same atom. For concreteness, we focus on ground-clock and clock-Rydberg Rabi
frequencies of $\Omega_\t{c} = 2 \pi \times 200\,\t{kHz}$ and $\Omega_\t{R} = 2
\pi \times 6\,\t{MHz}$, respectively, and we show operation fidelities on both
transitions exceeding 0.99 under magnetic fields of $B \gtrsim
200\,\t{G}$ and polarization impurities of $\chi \gtrsim 10^{-2}$. Additionally,
we analyze temperature effects on the clock transition (and refer to
Refs.~\cite{deLeseleuc2018, Madjarov2020} and Appendix~\ref{RydbergNoise} for consideration of such effects on the
Rydberg transition), finding that $T \lesssim 10\,\t{$\mu$K}$ is sufficient for
clock pulses with fidelity exceeding 0.99. Finally, we analyze single-beam Raman
gates for rotations of nuclear spin qubits and identify a ``magic'' linear
polarization angle where the pulse-induced light shift perfectly cancels the
nuclear Zeeman shift. We show the feasibility of purely $\sigma_x$ rotations
with $\Omega_\t{eff} \approx 2 \pi \times 1\,\t{MHz}$ at fidelities exceeding
$0.99$. All these conditions are readily available in current experiments.

We specifically considered ${}^{171}\t{Yb}$ to exploit its $I = 1/2$, built-in
nuclear spin qubits; however, other isotopes including ${}^{173}\t{Yb}$ and
${}^{87}\t{Sr}$ with larger $I$ offer similar opportunities albeit with
additional control fields required to isolate only two nuclear spin
states~\cite{Barnes2021}. Nevertheless, larger-$I$ isotopes offer unique
opportunities for $SU(N)$ physics~\cite{Gorshkov2010, Scazza2014} and
higher-dimensional computational spaces such as
qudecimals~\cite{Omanakuttan2021} that could be leveraged for robust
encoding~\cite{Albert2020}. In terms of the structure of $S$-series Rydberg
states for isotopes with $I > 1/2$, we expect a similar behavior where the
${}^3\t{S}_1$ $F_\t{max} = 1 + I$ is well-behaved since it is a unique
configuration of electron and nuclear spins ~\cite{Robicheaux2018,  Ding2018,
Robicheaux2019, Shi2021b}.

The \textit{omg} architecture discussed in this work uniquely enables new
opportunities for shelving-based readout~\cite{Covey2019, Barnes2021, Monz2016,
Erhard2021} as well as remote entanglement~\cite{Covey2019b,Huie2021}. However,
we note that other variants of this versatile \textit{omg} architecture offer
additional opportunities not discussed here. More generally, this platform holds
promise for programmable entanglement in atomic clocks~\cite{Gil2014,
Kessler2014, Kaubruegger2019}, quantum networking~\cite{Covey2019b, Huie2021},
and quantum computation~\cite{Daley2008, Gorshkov2009, Gorshkov2010, Cong2021}.
A similar \textit{omg} architecture has recently been
proposed~\cite{Allcock2021} and demonstrated~\cite{Yang2021} for trapped ions,
where the additional required primitive operations are already compatible with
existing large-scale systems. We believe the same is true for the neutral
AEA-based platform~\cite{Choi2021, Young2020, Burgers2021, Barnes2021,
Jenkins2021, Ma2021}.

\section*{Acknowledgments}
We thank Brett Merriman, Abhishek Desai, Ivaylo Madjarov, Hannes Bernien, Adam
Kaufman, and Jeff Thompson for helpful discussions. We acknowledge funding from
the NSF QLCI for Hybrid Quantum Architectures and Networks (NSF award 2016136),
the NSF PHY Division (NSF award 2112663), and the NSF Quantum Interconnects
Challenge for Transformational Advances in Quantum Systems (NSF award 2137642).
C. H. G. is supported in part by the AFOSR-MURI, grant number
FA9550-20-1-0323.

\textit{Note:} We became aware of another work considering Rydberg-mediated
gates for metastable ${}^3\t{P}_0$ nuclear spin qubits in ${}^{171}\t{Yb}$ while
completing this manuscript~\cite{Wu2022}.

\setcounter{section}{0}
\twocolumngrid

\renewcommand\thefigure{S\arabic{figure}}
\renewcommand\thetable{S\arabic{table}}
\setcounter{figure}{0}  

\renewcommand\appendixname{APPENDIX}
\appendix
\renewcommand\thesection{\Alph{section}}
\renewcommand\thesubsection{\arabic{subsection}}


\section{Multichannel quantum defect theory}\label{MQDT}
Previous studies of neutral Yb Rydberg levels have determined a multichannel
quantum defect theory (MQDT) representation of the energy level spectrum,
including perturbing levels of valence character (such as $6p^2$ or
$4f^{13}5d^26p$).  For the spin 0 isotopes of Yb, this provides a nearly
complete characterization of many symmetries of the Rydberg series in the energy
range extending to approximately 0.05 eV (12.1 THz) below the lowest ionization
threshold. However, for a nuclear spin $I=1/2$ isotope such as $^{171}$Yb, the
hyperfine splitting can couple different $J$ channels, and in particular the
hyperfine interaction causes a strong coupling between the $6sns \,{ }^1\t{S}_0$
and $6sns\,{ }^3\t{S}_1$ Rydberg series that gets very strong for Rydberg state
binding energies that are comparable to the hyperfine splitting in the Yb$^+$
ion.

\begin{figure}[t!]
	\centering
	\includegraphics[width=0.48\textwidth]{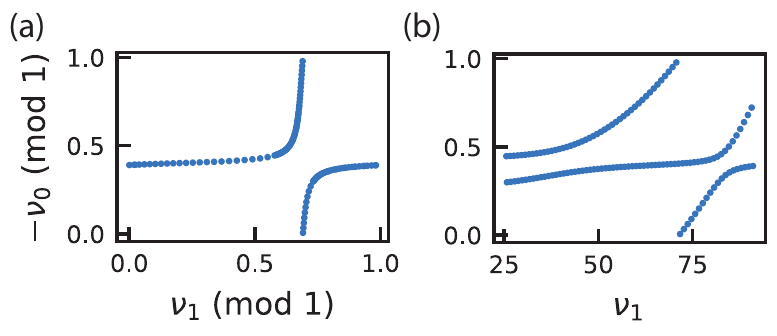}
    \caption{
        \textbf{Lu-Fano plots for the bound state quantum defects of the
        two-channel $F=1/2$ series}. Shown in (a) is the periodic version of the
        Lu-Fano plot~\cite{Lu1970}, with the quantum defect relative to the
        lower ($f_\t{c} = 0$) hyperfine ionization threshold on the $y$-axis,
        modulo 1, and the effective quantum number relative to the upper
        ($f_\t{c} = 1$) threshold on the $x$-axis, also modulo 1. In the
        approximation used here, namely with an energy-independent $K$-matrix as
        is written above, the plot is exactly periodic. The extent of the
        channel coupling is reflected in the strength of the avoided crossing
        near the center of the figure.  In (b) the plot shows the same bound
        levels, but without applying the (modulo 1) to the $x$-axis effective
        quantum number data. The approximately  horizontal branch is close to
        the $^3S_1$ quantum defect value, and from (b) it can be deduced that no
        significant level perturbations to that series should occur for $\nu_1$
        in the range 40 to 75.
        \label{FigureLuFano}
    }
\end{figure}

The basic theory that describes hyperfine-induced coupling of different
electronic angular momentum channels in an atom follows the basic ideas of the
frame transformation theory introduced into MQDT (FT-MQDT) by Fano, Lu, and
Lee~\cite{Fano1970, Lu1970, Lee1973}. The theory was adapted to the specific
context of hyperfine coupling by Sun and Lu~\cite{Sun1988, Sun1989} and extended
to heavier complex atoms by Robicheaux {\it et al.}~\cite{Robicheaux2018}. Our
implementation of the theory in the present context focuses on the two channels
that have singlet and triplet character mixed primarily by the hyperfine
splitting of the Yb$^+(6s_{1/2})$ ionic core.  We omit the closed subshell
$4f^{14}$ from our notation except in contexts where its open-shell excitations
arise.  In FT-MQDT, the key quantity to determine is the reaction matrix
$\underline{K}$, in an appropriate representation of the long-range channels.

For the $6sns$ Rydberg states of interest here, the only angular momentum
quantum numbers are the ionic core spin $s_\t{c}=1/2$, the Rydberg electron spin
$s=1/2$, and the nuclear spin, $I=1/2$ for $^{171}$Yb. The reaction matrix is
first determined for each value $J=S$ of the electronic angular momentum, which
is a good quantum number when neglecting the hyperfine interaction altogether.
The singlet quantum defect $\mu_0$ used here has been taken from
Ref.~\cite{Lehec2018}, while the triplet $\mu_1$ is taken from
Ref.~\cite{Wilson2019}. Specifically, the electronic reaction matrix is diagonal
in the singlet-triplet representation, i.e. $K_{SS'}=\delta_{SS'}\tan{\pi
\mu_S}$. Note that we have approximated the singlet and triplet quantum defects
as energy-independent, but this could easily be improved to obtain spectroscopic
accuracy for these calculations. When the nuclear spin Hilbert space is
included, this ``eigenchannel representation''~\cite{Aymar1996} of the reaction
matrix for the quantum number $F=1/2$ characterizing the total angular momentum
${\textbf{F}}={\textbf{I}}+{\textbf{S}}$ has the structure:
\begin{equation}
    \langle [I(s_\t{c} s)S]F M_F|\underline{K}|[I(s_\t{c} s)S']F M_F \rangle.
\end{equation}
The first step of the FT-MQDT is application of a straightforward recoupling
into a representation that includes the total angular momentum quantum number of
the core.  That is needed because the ionization thresholds depend on the {\it
ionic} core total angular momentum $f_\t{c}=0$ or 1, where $\textbf{f}_c =
\textbf{I} + \textbf{s}_\t{c}$. The recoupling coefficient looks like $\langle
(I s_\t{c})f_\t{c}|[I(s_\t{c} s)S] \rangle^{(F)}$, which is proportional to a
6-$j$ coefficient as in standard references. The resulting 2-channel FT-MQDT
K-matrix which can be viewed as energy-independent for sufficiently high Rydberg
states with $n \gtrsim 35$ is equal to:
\begin{equation}
    \underline{K}
        = \left(\begin{array}{cc}
             4.1088 & 1.6922 \\
             1.6922 & 2.1549 \\
        \end{array}\right)
,\end{equation}
where the first channel corresponds to the lower ionic hyperfine threshold
$f_\t{c} = 0$ and the second channel corresponds to the upper threshold $f_\t{c}
= 1$.  If we set the zero of our energy scale to the degeneracy-weighted average
of the two hyperfine thresholds, the two threshold energies $E_{f_\t{c}}$ are
given in terms of the hyperfine splitting $\Delta_\t{HFS} = 2 \pi \times
12.6428121\,\t{GHz}$ as $\{E_0 / h = -9.482109, E_1 / h = 3.160703\}\,\t{GHz}$.

At this point, bound state energies $E_n$ are determined by solving for roots of
the following equation:
\begin{equation}
  \det\left\{ {\underline K}+\tan{\pi {\underline \nu}} \right\}
        = 0
,\end{equation}
where the diagonal matrix ${\underline \nu}$ consists of effective quantum
numbers in the two channels, defined for energies below the lower threshold, by:
\begin{equation}
    \label{eqn: effective quantum numbers}
    \nu_{f_\t{c}}(E)
        = \sqrt{\frac{{\rm Ry}({}^{171}{\rm Yb})}{E_{f_\t{c}} - E}}. 
\end{equation}
Here, ${{\rm Ry}({}^{171}{\rm Yb})}$ is the Rydberg constant for this
electron-ion system, {\it i.e.} the infinite mass Rydberg constant multiplied by
the ratio between the reduced electron-${}^{171}{\rm Yb}^+$ mass and the bare
electron mass.

The resulting bound state Rydberg energy levels are displayed in the form of
Lu-Fano plots~\cite{Lu1970} in Figure~\ref{FigureLuFano}. These Lu-Fano plots
illustrate the behavior of the $F=1/2$ Rydberg series as the principal quantum
number increases. The energy levels with respect to the ground state are
obtained by inverting \eqref{eqn: effective quantum numbers} to calculate
$E_{f_\t{c}}$, and subsequently shifting them by the energy of the lower ionic
hyperfine threshold relative to the ground state. These values are plotted in
Figure~\ref{FigureRydbergHyperfine}(d). We note that near-degeneracies occur
between the two series in the region where they begin to diverge and then slip
by modulo 1 in Fig.~\ref{FigureLuFano}. It is thus best to avoid this regime,
which is why we focus on $n^*\approx55$.

It should be noted that the present 2-channel model of the $6sns$ Rydberg series
does not include some of the channels that can cause additional perturbations,
as have been studied in the literature.  See Figures 4 and 5 of
Ref.~\cite{Lehec2018}, for example, which shows that level perturbations such as
$4f^{14}6p^2$ and $4f^{13}5d6s6p$ occur for low principal quantum numbers below
about $n\approx 25$, but these are unlikely to occur for any of the Rydberg
series considered in the present study. Strictly speaking, the $F = 1/2$ Rydberg
series and Lu-Fano plot should include the $6snd^3\t{D}_1$ Rydberg series as
well, but our estimates suggest that the amplitude of mixing with the
$6sns^3\t{S}_1$ series is small and only of order $10^{-3}$, and for this reason
the $6snd$ series are not included in our MQDT model. Moreover, the
${}^3\t{D}_1$ and ${}^3\t{D}_2$ quantum defects are in the range 0.72-0.76 and
thus well separated from the $6sns$ levels of interest here.  Similarly, an
exact treatment of the $F=3/2$ series would include the coupling of $^3\t{S}_1$
states to ${}^1\t{D}_2,\ {}^3\t{D}_1,{\rm and}\ {}^3\t{D}_2$ series, but those
are also neglected here because the coupling is expected to be small for this
total angular momentum as well.

\section{The $F = 3/2$ ${}^3\t{S}_1$ Rydberg series}\label{3S1}
\subsection{Bare energies relative to $F = 1/2$}
The $F = 3/2$ ${}^3\t{S}_1$ Rydberg series is a simpler series to handle than
the $F = 1/2$ series due to the fact that it is a single channel converging to
the $f_\t{c} = 1$ ionization threshold. In order to calculate the energy levels,
however, we require the knowledge of the quantum defect of the ${}^3\t{S}_1$
Rydberg series. Due to a lack of experimental spectroscopic data for
${}^{171}\t{Yb}$, we draw upon available data for the bosonic ${}^{174}\t{Yb}$
isotope to deduce the quantum defect. In particular, the ${}^3\t{S}_1$ series
has been mapped out in Ref.~\cite{Wilson2019}. The energy levels for the
${}^{171}\t{Yb}$ $F = 3/2$ ${}^3\t{S}_1$ series are obtained by finding the
effective quantum numbers from the measured levels and applying them in
\eqref{eqn: effective quantum numbers}. Note that we use the $f_\t{c} = 1$
ionization threshold. This shows that the $F = 3/2$ and $F = 1/2$ series are
well-separated by at least $ \approx10\,\t{GHz}$ over the range of effective
quantum number $n^*$ shown in Figure~\ref{FigureRydbergHyperfine}(d), eventually
widening to the hyperfine splitting of the ionic core of $\Delta_\t{HFS} = 2 \pi
\times 12.6\,\t{GHz}$.

\subsection{g-factor of the $F = 3/2$ series}
Due to the simplicity of the single channel nature of this series, the
respective $S_\t{total}$, $J$, and $F$ angular momenta are well-defined, with
the caveats mentioned at the end of Appendix~\ref{MQDT}. This permits the use of
the standard result for calculating the g-factor for this series at low magnetic
fields. At low fields, the total angular momentum $\mathbf{F}$ precesses about
the applied field. Thus, we aim to write
\begin{equation}
	E_\t{Z}
        = -\langle \boldsymbol{\mu} \cdot \mathbf{B} \rangle
        = g_F m_F \mu_B B,
\end{equation}
where $g_F$ is the g-factor of interest.

The magnetic moment depends on the total spin of the electrons $\mathbf{S}$ and
the nuclear spin $\mathbf{I}$. Since $\mathbf{F} = \mathbf{S} + \mathbf{I}$, we
can project the respective angular momenta onto $F$ to evaluate the matrix
element:
\begin{equation}
	\langle \mathbf{A} \rangle
        = \frac{
            \langle \mathbf{A} \cdot \mathbf{F} \rangle
        }{
            F (F + 1)
        }
            \langle \mathbf{F} \rangle,
\end{equation}
The dot product can be evaluated easily as
\begin{gather}
	\langle \mathbf{S} \cdot \mathbf{F} \rangle
        = \frac{\hbar m_F}{2 F (F + 1)}
            \left[F (F + 1) + S (S + 1) - I (I + 1)\right],
    \\
	\langle \mathbf{I} \cdot \mathbf{F} \rangle
        = \frac{\hbar m_F}{2 F (F + 1)}
            \left[F (F + 1) + I (I + 1) - S (S + 1)\right].
\end{gather}
Packaging everything together gives
\begin{align}
    \begin{aligned}
        g_F
            &= g_S \frac{F (F + 1) + S (S + 1) - I (I + 1)}{2 F (F + 1)}
            \\&\phantom{=}
            - g_I \frac{\mu_N}{\mu_B}
                \frac{F (F + 1) - S (S + 1) + I (I + 1)}{2 F (F + 1)}
    \end{aligned}
\end{align}
With $g_S = 2$, $g_I = 0.4919$, $F = 3/2$, $S = 1$, $I = 1/2$, the g-factor
evaluates to $1.9\,\t{MHz/G}$.

\subsection{Diamagnetic shift of the Rydberg series}
As mentioned in the main text, the Rydberg states experience an additional
diamagnetic shift in its energy due to a magnetic field. The diamagnetic
Hamiltonian, given by
\begin{equation}
	H_\t{DM}
        = \frac{1}{8 m_e} \abs{\mathbf{d} \times \mathbf{B}}^2,
\end{equation}
arises from the term quadratic in the vector potential $\mathbf{A}$ in the
Hamiltonian for a charged particle in an external electromagnetic field. This
quadratic term is typically neglected in comparison to the linear term
($\mathbf{A} \cdot \mathbf{p}$), which is responsible for the linear Zeeman
effect. However, due to the scaling of $\mathbf{d}$ as $n^2$ for Rydberg atoms,
we anticipate that the quadratic term is comparable or even larger than the
linear term. Thus, it is important to explicitly determine the energy shift due
to the diamagnetic interaction.

To calculate the diamagnetic shift, it will be fruitful to expose the angular
dependence of the Hamiltonian by writing it in terms of spherical harmonics
$Y_{lm}(\theta,\phi)$. Since the cross product squared yields a factor of
$\sin^2\theta = 1 - \cos^2\theta$, we can rewrite it as
\begin{align}
	\Delta E_\t{DM}
        &= \frac{e^2 B^2}{8 m_e} \langle r^2 \sin^2\theta \rangle,
        \\
	    &= \frac{e^2 B^2}{8 m_e} \frac{4 \sqrt{\pi}}{3}
            \left\langle
                r^2 \left(Y_{00} - \frac{1}{\sqrt{5}} Y_{20} \right)
            \right\rangle.
\end{align}
An application of the Wigner-Eckart theorem reduces the problem to calculating
the reduced matrix element of the $r^2$ operator and the factors arising from
the angular dependence. The former can be dealt with using a variety of
numerical tools developed in recent years to calculate matrix elements of
Rydberg states. In particular, we utilize the ``Alkali.ne Rydberg Calculator''
(ARC) 3.0 package~\cite{Robertson2021} as the code has been expanded recently to
support calculations for AEAs. For the angular-dependent factors, we find that
for the ${}^3\t{S}_1$, $F = 3/2$ manifold, only $Y_{00}$ contributes a non-zero
value. Moreover, it is independent of the $m_F$ values. It follows that the four
Zeeman states experience the same diamagnetic shift which scales as
\begin{equation}
	\Delta E_\t{DM} / h
        = 2.4 ~\text{kHz/G\textsuperscript{2}}.
\end{equation}
Comparing with the linear Zeeman shift of $1.9\,\t{MHz/G}$, we see
that the two shifts become comparable at $\sim 800\,\t{G}$. Thus, we may neglect
the diamagnetic shifts for most purposes. In any case, the diamagnetic shift
does not affect the energy selectivity due to the equal shifts of all $m_F$
states.

\subsection{Hyperfine mixing between Rydberg series}
We address the possibility of hyperfine mixing within the ${}^3S_1$ Ryberg
manifold by diagonalizing the full Zeeman Hamiltonian for the Rydberg atom,
treating the nucleus, the $6s$ core electron, and the Rydberg electron as
separate entities. The basis of choice is the hyperfine basis
$\ket{[(Is_\t{c})f_\t{c} s]F M_F}$. We find that the Zeeman shift is linear for
the $m_F = \pm \frac{1}{2}$ states in the two series, up to 1000 G, indicating
that there is no significant mixing between the Rydberg series.

Another possible mixing channel is the diamagnetic coupling between the
${}^3S_1$ and ${}^3D_J$ manifolds. This arises from the $Y_{20}$ term in the
diamagnetic Hamiltonian. We assume that the coupling is significant when
$\abs{\langle {}^3S_1| H_\t{DM} |{}^3D_J
\rangle}/\abs{E({}^3D_J)-E({}^3S_1)}\gtrsim 0.1$, corresponding to $\approx10$\%
amplitude admixture. To get an order of magnitude estimate, we neglect the
angular dependency in $\langle r^2 \sin^2 \theta \rangle$ by taking $\langle r^2
\rangle \sim 5n^4/2$, effectively setting an upper bound for the matrix element,
and use $\abs{E({}^3D_J)-E({}^3S_1)} \sim 0.3/n^3$ (atomic units). For
$n^*\approx55$, the 10\% amplitude admixture occurs at $B\approx600$ G,
rendering this effect negligible at $\approx200$ G.

\section{Hyperfine mixing in the ``clock'' state}\label{HFMix}
For the bosonic species of AEAs, the clock transition is typically
doubly-forbidden as it is a $J = 0$ to $J' = 0$ transition, with $\Delta S = 1$.
On the other hand, the fermionic species has a weak admixture of the
${}^3\t{P}_0$ clock state with the ${}^1\t{P}_1$ state arising from the
hyperfine mixing of states with the same $F$. This small ${}^1\t{P}_1$ character
in the clock state enables a non-zero electric dipole coupling between the clock
and ground states.

Although the hyperfine mixing allows us to drive the transition between the
ground and clock states at large Rabi frequencies ($\sim 200\,\t{kHz}$ as stated
in the main text), the hyperfine mixing complicates the Zeeman effect
experienced by the clock hyperfine sublevels in the presence of a magnetic
field. The full Zeeman effect is described by the total Hamiltonian
\begin{equation}\label{eqn: total Zeeman Hamiltonian}
	H_\t{total}
        = H_Z + H_A + H_Q,
\end{equation}
where we have the usual Zeeman Hamiltonian
\begin{equation}\label{eqn: Zeeman Hamiltonian}
	H_Z
        = -\boldsymbol{\mu} \cdot \mathbf{B},
\end{equation}       
and the corrections from the hyperfine and quadrupole effects
\begin{equation}
	H_A + H_Q
        = A \mathbf{I} \cdot \mathbf{J}
        + Q \frac{
            \frac{3}{2} \mathbf{I}
                \cdot \mathbf{J} (2 \mathbf{I} \cdot \mathbf{J} + 1)
            - I J (I + 1) (J + 1)
        }{
            2 I J (2 I - 1) (2 J - 1)
        }.
\end{equation}
We will need to diagonalize \eqref{eqn: total Zeeman Hamiltonian} in order to
describe the Zeeman effect across all values of the applied magnetic field. We
adopt the methods and convention of ~\cite{Boyd2007} to calculate the Zeeman map
of the clock state across a large range of magnetic field values. Accordingly,
the Zeeman Hamiltonian of \eqref{eqn: Zeeman Hamiltonian} is written as
\begin{equation}
		H_Z = (g_s S_z + g_l L_z - g_I I_z)\mu_0 B,
\end{equation}
where $g_s \approx 2$, $g_l = 1$, $g_I = \frac{\mu_I}{\mu_B |I|}$ are the $g$
factors of the electron spin, orbital angular momentum, and nuclear spin
respectively; and $\mu_0 = \mu_B / h$ is the Bohr magneton in units of Hz/T. The
angular momentum operators here are dimensionless. The quadrupole Hamiltonian
can be dropped as $Q = 0$ for $I = 1/2$~\cite{Berends1992}. Thus, the only
correction that we need to include is $H_A$.

For the ${}^1\t{S}_0$ ground state, it experiences only a linear Zeeman shift
due to the fact that $\mathbf{J} = 0$, hence there is no hyperfine correction.
Thus, the energy shift (in units of Hz) is
\begin{equation}
    \Delta \nu({}^1\t{S}_0, m_F)
        = -g_I m_F \mu_0 B.
\end{equation} 

For the ${}^3\t{P}_0$ clock state, the hyperfine mixing between the
${}^3\t{P}_0$ and ${}^3\t{P}_1$ states leads to a Breit-Rabi expression given by
\begin{align}
    \begin{aligned}
        \nu({}^3P_0, m_F)
            &= \frac{1}{2} \left(\nu({}^3\t{P}_0) + \nu({}^3\t{P}_1)\right)
            + \frac{1}{2} \left(\nu({}^3\t{P}_0) - \nu({}^3\t{P}_1)\right)
        \\&\phantom{=}
        \times \sqrt{
            1 + 4 \frac{
                \sum_{F'} \alpha^2
                |\bra{{}^3\t{P}_0^0, F} H_Z \ket{{}^3\t{P}_1, F'}|^2
            }{
                \left(\nu({}^3\t{P}_0) - \nu({}^3\t{P}_1)\right)^2
            }
        }
    \end{aligned},
\end{align}
where
\begin{gather}
	\begin{aligned}
        \nu({}^3\t{P}_0)
            &= \nu({}^3\t{P}_0^0) + \bra{{}^3\t{P}_0^0} H_Z \ket{{}^3\t{P}_0^0}
            \\&\phantom{=}
		    + 2 (\alpha_0 \alpha - \beta_0 \beta)
                \bra{{}^3\t{P}_1^0, F = I} H_Z \ket{{}^3\t{P}_0^0},
	\end{aligned}
    \\
	\begin{aligned}
        \nu({}^3\t{P}_1)
            = \nu({}^3\t{P}_1^0)
            + \sum_{F'}&\bigg(
                \alpha^2 \bra{{}^3\t{P}_1^0, F'} H_Z \ket{{}^3\t{P}_1^0, F'}
                \\&
                + \beta^2 \bra{{}^1\t{P}_1^0, F'} H_Z \ket{{}^1\t{P}_1^0, F'}
            \bigg).
	\end{aligned}
\end{gather}
The matrix elements are taken between states of pure $LS$ nature, as denoted by
the superscript 0. The constants \{$\alpha,\, \beta$\} and \{$\alpha_0,\,
\beta_0$\} are known as the intermediate coupling and hyperfine mixing
coefficients as they characterize the extent of the admixture of the atomic
states:
\begin{gather}
    \ket{{}^3\t{P}_0}
        = \ket{{}^3\t{P}_0^0},
    \\
    \ket{{}^3\t{P}_1}
        = \alpha \ket{{}^3\t{P}_1^0} + \beta \ket{{}^1\t{P}_1^0},
\end{gather}
and
\begin{equation}
    \begin{aligned}
        \ket{{}^3P_0, I, F}
            &= \ket{{}^3\t{P}_0^0}
            + (\alpha_0 \alpha - \beta_0 \beta) \ket{{}^3\t{P}_1^0}
        \\&\phantom{=}
        + (\alpha_0 \beta + \beta_0 \alpha) \ket{{}^1\t{P}_1^0}.
    \end{aligned}
\end{equation}
Most importantly, these coefficients are related to experimentally measurable
quantities:
\begin{gather}
    \tau({}^3\t{P}_1)
        = \left(\frac{\nu({}^1\t{P}_1)}{\nu(^3P_1)}\right)^3
            \frac{\alpha^2}{\beta^2}\tau(^1P1);
    \\
    \tau({}^3\t{P}_0)
        = \left(\frac{\nu({}^3\t{P}_1)}{\nu(^3P_0)}\right)^3
            \frac{\beta^2}{(\alpha_0 \beta + \beta_0 \alpha)^2}
            \tau({}^3\t{P}_1);
    \\
	\delta g
        = (\alpha_0 \alpha - \beta_0 \beta) \sqrt{\frac{8}{3 I (I + 1)}},
\end{gather}
where $\tau$ is the lifetime of the state, and $\delta g$ is the differential
g-factor for the clock state, such that $g_I({}^3\t{P}_0) = g_I + \delta g$ at
weak magnetic fields. These expressions can be used to estimate the values of
the coupling constants, which are summarized in Table~\ref{table:Breit-Rabi
parameters}.

\begin{table}[t!]
    \renewcommand*{\arraystretch}{1.3}
	\caption{
        Table of parameters for ${}^{171}\t{Yb}$. Parameters with $^\dagger$ are
        taken from~\cite{Lemke2012}.
        \label{table:Breit-Rabi parameters}
    }
	\begin{ruledtabular}
		\begin{tabular}{cc}
			Parameter & Value\\
            \hline
			$\abs{\alpha}^\dagger$  & 0.996\\
			$\abs{\beta}^\dagger$  & 0.125\\
			$\delta g^\dagger$  & 2.73$\times 10^{-4}$\\
			$\abs{\alpha_0}$ & 1.41$\times 10^{-4}$\\
			$\abs{\beta_0}$ & 3.33$\times 10^{-5}$\\
		\end{tabular}
	\end{ruledtabular}
\end{table}

\section{Numerical simulation of multilevel dynamics}\label{MultilevelSim}
\subsection{Method overview}
We employ a numerical model to analyze the dynamics of the clock and Rydberg
multilevel systems. For a general system of $n$ states composing the basis
$S = \{\ket{1}, \dots, \ket{n}\}$ with energies $\hbar \times \{\omega_1, \dots,
\omega_n\}$, we write the total, time-dependent state $\ket{\psi(t)}$ as
\begin{equation}
    \label{eq:state-definition}
    \ket{\psi(t)}
        = \sum\limits_{k = 1}^n a_k(t) e^{-i \omega_k t} \ket{k},
\end{equation}
where its ``free-evolving'' components have been explicitly divided out from the
amplitudes $a_1, \dots, a_n$. This choice is convenient for the later
computation of phases discussed in Appendix~\ref{PhaseAccrual}. In this frame,
the Hamiltonian for the system in the presence of a drive of strength $\Omega$
and frequency $\omega$ has only off-diagonal components,
\begin{equation}
    \label{eq:single-atom-H}
    \hat{H}(t)
        = \hbar \sum\limits_{b = 1}^n \sum\limits_{a < b}
            \frac{\Omega}{2} g_a^b(\chi, q)
                e^{i (\omega - \omega_0 - \tld{\omega}_a^b) t}
            \ketbra{b}{a}
        + \t{H.c.},
\end{equation}
where the usual rotating wave approximation comparing $\omega$ to some chosen
reference energy $\omega_0$ (e.g. the difference in mean energies of the ground
and clock or clock and Rydberg manifolds) has been used, and $\tld{\omega}_a^b$
is the energy of the $a \leftrightarrow b$ transition relative to it. We also
consider a transition-dependent factor $g_a^b(\chi, q)$ modulating the
``principal'' drive strength $\Omega$ of the targeted transition. $g_a^b(\chi,
q)$ provides the correct couplings for specific polarizations $q \in \{0, \pm
1\}$ of the drive field, with additional weighting for impurities $\chi$ therein
as well as Clebsch-Gordan coefficients, as discussed in the main text. In
general, $\Omega$ and $\omega$ may be time-dependent as well to account for
intensity and/or phase noise, respectively (see Appendix~\ref{PhaseNoise}), in
which case we take $\omega t \to \phi(t) = \int_0^t \omega(t') \dd{t'}$.

\begin{table}[t!]
    \renewcommand*{\arraystretch}{1.2}
    \caption{
        Table of Clebsch-Gordan weighting factors for all transitions of
        interest in this work in the presence of a $\sigma^+$ ($q = 1$) drive,
        according to Eq.~(\ref{eq:CG}).
        \label{table:CG}
    }
    \begin{tabular}{
            >{\centering\arraybackslash}p{0.05\textwidth}
            >{\centering\arraybackslash}p{0.05\textwidth}
            >{\centering\arraybackslash}p{0.05\textwidth}
            >{\centering\arraybackslash}p{0.05\textwidth}
            >{\centering\arraybackslash}p{0.225\textwidth}
        }
        \midrule\midrule
        $F^a$ & $m_F^a$ & $F^b$ & $m_F^b$ & $W(F^a, m_F^a, F^b, m_F^b, 1)$ \\
        \midrule
        \multicolumn{5}{c}{Ground-clock (${}^1\t{S}_0 \leftrightarrow {}^3\t{P}_0$)} \\
        \midrule
        \multirow{4}{*}{$1/2$} & $+1/2$ & \multirow{4}{*}{$1/2$} & $+1/2$ & $\sqrt{1/2}$ \\
                               & $+1/2$ &                        & $-1/2$ & $1$ \\
                               & $-1/2$ &                        & $+1/2$ & $1$ \\
                               & $-1/2$ &                        & $-1/2$ & $\sqrt{1/2}$ \\
        \midrule
        \multicolumn{5}{c}{Clock-Rydberg (${}^3\t{P}_0 \leftrightarrow {}^3\t{S}_1$)} \\
        \midrule
        \multirow{6}{*}{$1/2$} & $+1/2$ & \multirow{6}{*}{$3/2$} & $+3/2$ & $1$ \\
                               & $+1/2$ &                        & $+1/2$ & $\sqrt{2/3}$ \\
                               & $+1/2$ &                        & $-1/2$ & $\sqrt{1/3}$ \\
                               & $-1/2$ &                        & $+1/2$ & $\sqrt{1/3}$ \\
                               & $-1/2$ &                        & $-1/2$ & $\sqrt{2/3}$ \\
                               & $-1/2$ &                        & $-3/2$ & $1$ \\
        \midrule\midrule
    \end{tabular}
\end{table}

Expanding further on the transition-dependent drive strength modulation factor
$g_a^b(\chi, q)$, we formally define this quantity in terms of two
distinct parts,
\begin{equation}
    \label{eq:drive-modulation}
    \begin{split}
        g_a^b(\chi, q)
            &= \rho(\chi; m_F^{a}, m_F^{b}, q)
            \\&\phantom{=}
            \times W(F^{a}, m_F^{a}, F^{b}, m_F^{b}, q)
    .\end{split}
\end{equation}
The first, $\rho(\chi; m_F^{a}, m_F^{b}, q)$, accounts for effects due to
polarization impurity in terms of the parameter $\chi$ introduced in the main
text. With $q$ held fixed for a given drive polarization, the corresponding
weighting factor is $\sqrt{1 - \chi}$ for transitions satisfying $m_F^{b} -
m_F^{a} = q$, while for all other, ``parasitic'' transitions, the factor is
$\sqrt{\chi / 2}$ to conserve total power in the drive across all three possible
polarizations,
\begin{equation}
    \label{eq:polarization-weight}
    \rho(\chi; m_F^{a}, m_F^{b}, q)
        = \begin{cases}
            \sqrt{1 - \chi}
                & \t{if}~ m_F^{b} - m_F^{a} = q
            \\
            \sqrt{\chi / 2}
                & \t{otherwise}
        \end{cases}
.\end{equation}
The second, $W(F^{a}, m_F^{a}, F^{b}, m_F^{b}, q)$, imposes weighting by
Clebsch-Gordan coefficients and dipole selection rules on all non-principal
transitions, normalized to that for the targeted transition. This factor is
conveniently defined in terms of the usual Wigner 3-$j$ symbols,
\begin{equation}
    \label{eq:CG}
    W(F^{a}, m_F^{a}, F^{b}, m_F^{b}, q)
        = \frac{
            \begin{pmatrix}
                F^{b} & 1 & F^{a}
                \\
                m_F^{b} & m_F^{a} - m_F^{b} & -m_F^{a}
            \end{pmatrix}
        }{
            \begin{pmatrix}
                F^{b} & 1 & F^{a}
                \\
                \bar{m}_F^{b} & -q & -\bar{m}_F^{a}
            \end{pmatrix}
        }
\end{equation}
where $\bar{m}_F^a$ and $\bar{m}_F^b = \bar{m}_F^a + q$ are the quantum numbers
of the principal transition. The values of this function used for our
calculations are shown in Table~\ref{table:CG}.

With these definitions, we include as an example the form of the Hamiltonian for
the six-level clock-Rydberg manifold, subject to a $\sigma^+$ drive on resonance
with the $\ket{\uparrow_\t{m}} \leftrightarrow \ket{r}$ transition:
\begin{widetext}
	\renewcommand*{\arraystretch}{1.65}
	\begin{equation}
		\hat{H}_{\t{c} \leftrightarrow \t{R}}
		= \hbar \frac{\Omega}{2} \begin{bmatrix}
			0 & 0 & \t{H.c.} & \t{H.c.} & \t{H.c.} & 0
			\\
			0 & 0 & 0 & \t{H.c.} & \t{H.c.} & \t{H.c.}
			\\
			\sqrt{\frac{\chi}{2}}
                w\left(-\frac{1}{2}, -\frac{3}{2}\right)
                e^{i (3 \Delta + \delta) t}
			& 0 & 0 & 0 & 0 & 0
			\\
			\sqrt{\frac{\chi}{2}}
                w\left(-\frac{1}{2}, -\frac{1}{2}\right)
                e^{i (2 \Delta + \delta) t}
			& \sqrt{\frac{\chi}{2}}
                w\left(+\frac{1}{2}, -\frac{1}{2}\right)
                e^{i 2 \Delta t}
			& 0 & 0 & 0 & 0
			\\
			\sqrt{1 - \chi}
                w\left(-\frac{1}{2}, +\frac{1}{2}\right)
                e^{i (\Delta + \delta) t}
			& \sqrt{\frac{\chi}{2}}
                w\left(+\frac{1}{2}, +\frac{1}{2}\right)
                e^{i \Delta t}
			& 0 & 0 & 0 & 0
			\\
			0
			& \sqrt{1 - \chi}
			& 0 & 0 & 0 & 0
		\end{bmatrix}
		\begin{matrix}
			\ket{\downarrow_\t{m}}
			\\
			\ket{\uparrow_\t{m}}
			\\
			\ket{r_\Downarrow}
			\\
			\ket{r_\downarrow}
			\\
			\ket{r_\uparrow}
			\\
			\ket{r}
		\end{matrix}
		.\end{equation}
\end{widetext}
Here, we write the six-state basis for the clock-Rydberg manifold as
$\{\ket{\downarrow_\t{m}},\, \ket{\uparrow_\t{m}},\, \ket{r_\Downarrow},\,
\ket{r_\downarrow},\, \ket{r_\uparrow},\, \ket{r}\}$, where the Rydberg states
$\ket{r_X}$ are ordered by their $m_F$ values. For brevity, we also use
$w(m_F^a, m_F^b) \equiv W(1/2, m_F^a, 3/2, m_F^b)$ and define $\Delta$ and
$\delta$ as the differences in energy (up to a factor of $\hbar$) between the
adjacent $m_F$ states in the Rydberg and clock manifolds, respectively.

For the multi-atom case of the clock-Rydberg transition, we generate the
appropriate Hamiltonian for $N$ atoms in the product-state basis $\mathbb{S} =
S^N$ using the single-atom form in
Eq.~(\ref{eq:single-atom-H}):
\begin{equation}
    \begin{split}
        \hat{\mathbb{H}}_N(t)
            &= \sum\limits_{k = 1}^N
                \hat{\mathbb{I}}^{\otimes^{k - 1}}
                \otimes \hat{H}_k
                \otimes \hat{\mathbb{I}}^{\otimes^{N - k}}
            \\&\phantom{=}
            + \sum\limits_{\ket{A}, \ket{B} \in \mathbb{S}}
                V_{A, B} \ketbra{A}{B}
    \end{split}
\end{equation}
where $\hat{\mathbb{I}}$ is the $n \times n$ identity operator for a single
atom, $\hat{H}_k$ is the single-atom Hamiltonian for the $k$-th constituent, and
$\otimes$ denotes the Kronecker product. $V_{A, B}$ encodes interactions at the
atom-atom level between the $N$-atom states $\ket{A}$ and $\ket{B}$ including,
for instance, the $U_\t{VdW}$ Rydberg interaction.

Numerical simulation is accomplished using the standard fourth-order Runge-Kutta
integration scheme~\cite{Press2007} for the Schr\''odinger equation.
We define the grid of discretized times $t^k = k \dd{t},~ k = 0, \dots, N_t$
over which the state vector is integrated using the time-discretized Hamiltonian
$\hat{H}^k = \hat{H}(t^k)$ for $\dd{t} \ll 2 \pi / \Omega$ suitably short and
$N_t \dd{t}$ appropriately long.

\subsection{Magnetic field noise}
We are additionally interested in analyzing the effect of magnetic field noise
on the atomic dynamics. We first note that fluctuations should occur over time
scales corresponding to $\lesssim \t{kHz}$ frequencies due to large inductances
expected in coils found in realistic experimental apparatuses. Thus we can
assume that the field noise is slow compared to our laser pulses, and hence we
consider a field that varies only on a shot-to-shot basis. To simulate this, we
average the time evolution of the state vector over a series of $N$ trials (we
use $N = 30$ in our calculations), for each of which the magnetic field strength
$B$ is sampled from a Gaussian distribution with standard deviation $1\,\t{mG}$
and variable mean value held fixed for all trials. We choose the standard
deviation as a good approximation to the Johnson white noise found in servos
that are typically used to control the current in magnetic
coils~\cite{Moses2015, Covey2016}.

We consider magnetic field noise in this way for the analyses of both the
ground-clock and clock-Rydberg dynamics. We find that in both cases the effect
of this noise is negligible, and in the latter it is indiscernible. We therefore
only include it in this work for the ground-clock dynamics. The main effect of
this noise, as stated in the main text, is to reduce the coherence time of the
nuclear spin qubits to $T_2^* \gtrsim 1\,\t{s}$. However, this effect can be
mitigated by e.g. dynamical decoupling.

\subsection{Relative phase accrual on a qubit}\label{PhaseAccrual}
Since we calculate the full evolution of the state vector, the integration
scheme described above may also be used to find the relative phase accrued
between two basis states over some time interval. Given the calculated
time-dependent state vector $\ket{\psi^k} = \ket{\psi(t^k)}$, it is
straightforward to find the relative phase between two components $\ket{a}$ and
$\ket{b}$ of $\ket{\psi^k}$ as
\begin{equation}
    \label{eq:phase-accrual}
    \Delta\varphi_{a,b}^k
        = \arg\left(\frac{\braket{a}{\psi^k}}{\braket{b}{\psi^k}}\right)
.\end{equation}
We note here that, recalling Eq.~(\ref{eq:state-definition}), the free-evolving
components of the phase have already been explicitly removed, and hence
Eq.~(\ref{eq:phase-accrual}) gives the accrued phase due only to externally
applied drives to the dynamics.

For use in our numerical analysis of both the clock and Rydberg transitions, we
are interested in calculating this relative phase between the two $m_F$ states
of the ground (clock)-state manifold after an effective $2 \pi$-pulse has been
applied on the ground-clock (clock-Rydberg) transition. While the dynamics
governing the value of this phase are in general complicated for the systems
featured in this work, it is useful to consider the limit of strong magnetic
field and small polarization impurity. In this limit, there are essentially no
undesirable couplings, and hence both transitions simplify to a two-level system
(states $\ket{g}, \ket{e}$ representing one state of a qubit and its
corresponding excited state) undergoing Rabi oscillations with dressing from a
third, uncoupled spectator state $\ket{s}$ (representing the other state of the
qubit). We model the time dependence of the total state as
\begin{equation}
	\begin{split}
		\ket{\zeta(t)}
            &= \cos\left( \frac{\theta_0}{2} \right)
                \left[
                    \cos\left( \frac{\Omega}{2} t \right)
                        \ket{g}
                    + \sin\left( \frac{\Omega}{2} t \right) e^{i \gamma(t)}
                        \ket{e}
                \right]
            \\&\phantom{=}
            + \sin\left( \frac{\theta_0}{2} \right) e^{i \varphi_0} \ket{s}
	\end{split}
\end{equation}
where $\Omega$ is the Rabi frequency (defined in terms of oscillations in
probability, not amplitude), and $\gamma(t)$ depends on the polarization and
detuning of the drive. The constants $\theta_0$ and $\varphi_0$ describe the
initial state dressing, and we note that it is necessary to have $0 < \theta_0 <
\pi$ (i.e. to have non-zero initial population in both $\ket{g}$ and $\ket{s}$)
in order for the desired relative phase to be well defined. For the targeted
case of a resonant drive in this work, we also take $\gamma(t) = 0$. From this,
it is easily seen that at the targeted $2 \pi$ time $\tau_{2 \pi} = 2 \pi /
\Omega$, the relative phase accrued between the ground and spectator states over
the duration of the drive is invariably $\pi$ for all $\theta_0,
\varphi_0$,
\begin{equation}
	\ket{\zeta(\tau_{2 \pi})}
        = -\left[
            \cos\left( \frac{\theta_0}{2} \right) \ket{g}
            + \sin\left( \frac{\theta_0}{2} \right)
                e^{i (\varphi_0 + \pi)} \ket{s}
        \right]
.\end{equation}
We note that, as seen in Figs.~\ref{FigureRydbergDrive} and
\ref{FigureClockDrive}, the numerical calculations agree well with this expected
behavior.

\subsection{Modeling the single-beam Raman transitions}
In order to show the existence of the ``magic angle'' for the single-beam Raman
transitions between adjacent ``m''-qubit states as in Fig.~\ref{FigureRaman}, we
now move to a model featuring an atom in the presence of a driving field that
has well-defined polarization. In this model, we consider a linearly polarized
plane wave incident on an atom with $k$-vector perpendicular to a surrounding
magnetic field. Rather than its impurity $\chi$, we parameterize the wave's
polarization by modified Stokes parameters $\theta$, the angle between the
electric and magnetic fields. Computationally, this amounts to replacing the
weighting factor $\rho$ defined in Eq.~(\ref{eq:polarization-weight}) with
another formulation $\tld{\rho}$ to be derived below.

We considering a cylindrically symmetric system (spanned by orthogonal unit
vectors $\hat{r}$, $\hat{\varphi}$, and $\hat{z}$ in the typical fashion) with
magnetic field oriented along the $z$-axis. The incident plane wave is then
defined to have $k$-vector pointed along $\hat{r}$ and polarization vector
\begin{equation}
    \hat{\varepsilon}(\theta)
        = \hat{\varphi} \sin\theta + \hat{z} \cos\theta
.\end{equation}
Next, we define an additional set of orthogonal unit vectors $\hat{\epsilon}_q$
to describe the space of possible ways that a classical dipole moment may
rotate,
\begin{align}
    \hat{\epsilon}_{\pm 1}
        &= \frac{1}{\sqrt{2}} \big( \hat{\varphi} \pm i \hat{r} \big)
    ,&
    \hat{\epsilon}_0
        &= \hat{z}
,\end{align}
where the first two correspond to right-hand (parallel to $\hat{z}$) and
left-hand (anti-parallel to $\hat{z}$) rotation about the $z$-axis, associated
with $\sigma^\pm$ transitions, and the last to simple oscillation on the axis,
associated with the $\pi$ transition. Without loss of generality we may
associate $\hat{\epsilon}_{+1}$ with the $\sigma^+$ transition specifically, and
find the appropriate form for $\tld{\rho}$ as
\begin{align}
    \tld{\rho}(\theta; m_F^a, m_F^b)
        &= \hat{\epsilon}_{m_F^b - m_F^a}^*
            \cdot \hat{\varepsilon}(\theta)
        \\
        &= \begin{cases}
            \frac{1}{\sqrt{2}} \sin\theta
                & \t{if}~ |m_F^b - m_F^a| = 1
            \\
            \cos\theta
                & \t{if}~ m_F^b - m_F^a = 0
        \end{cases}
.\end{align}
This modified polarization weight $\tld{\rho}$ is then inserted into
Eq.~(\ref{eq:drive-modulation}), replacing each instance of $\rho$.

To produce the results shown in Fig.~\ref{FigureRaman}(c) and (d), we consider
driving the Raman transition specifically between the two ``m'' qubits, using
the ${}^3\t{D}_1 ~ F = 1/2,\, m_F = -1/2$ state as the intermediary, although
this system is easily mapped to cases where the use of a ${}^3\t{S}_1$
intermediary may be desirable, or when driving the ``g'' qubits (for equivalent
$F$ and $m_F$ quantum numbers of the intermediary state), as shown in
Fig.~\ref{OMG}(a). In our simulations, we use a drive strength corresponding to
$\Omega_V = 2 \pi \times 20\,\t{MHz}$ when the angle $\theta$ between the
driving electric and surrounding magnetic fields is zero. Estimating the reduced
dipole matrix element for the ${}^3\t{P}_0 \leftrightarrow {}^3\t{D}_1$
transition from a mixture of past determinations \cite{Beloy2012}, said drive
strength requires approximately $2\,\t{mW}$ of power in a beam of $1\,\t{mm}$ waist radius. Of course, the required power will be much lower for targeted Raman gates with a tightly-focused beam.

The subsequent analysis of the simulated dynamics of this system is identical to
that for the clock and Rydberg transitions above -- we evolve the system
initialized to the $\ket{\downarrow_\t{m}}$ clock state using a Hamiltonian of
the form given by Eq.~(\ref{eq:single-atom-H}) and calculate $\pi$-pulse fidelity
based on the resulting Rabi oscillations of the target $\ket{\uparrow_\t{m}}$
state -- with the exception that here the effective Rabi frequency
$\Omega_\t{eff}$ is also computed. Due to programmatic considerations, this is
done in two ways. Specifically, we find that conditions corresponding to large
regions of the considered parameter space give dynamics featuring high-frequency
probability oscillations of sufficient amplitude as to make the determination of
the effective $2 \pi$ time (and by extension the effective Rabi frequency)
difficult using only the time-domain oscillations. To combat this effect, we
compute $\Omega_\t{eff}$ by two methods: The first-maximum (FM) method is to
simply find the time corresponding to the first local maximum in the probability
oscillations of the initial state and invert it to find the corresponding
frequency. The second method (FT) is to find the frequency as the
lowest-frequency component of the Fourier transform of either the target or
initial states. The FM method is cheap to compute with good precision, but
strongly affected by the aforementioned problem with high-frequency
oscillations. On the other hand, the FT method escapes this problem, but
requires one to simulate the dynamics out to longer times in order to give good
resolution at low frequencies. For each set of conditions, $\Omega_\t{eff}$ is
obtained via both methods; if the two results agree to within $5\%$ of the FT
value, then the FM result is preferred, otherwise the FT. We find that
simulating to $25\,\t{$\mu$s}$ gives good low-frequency resolution under the
range of conditions considered.

\begin{figure}[t!]
	\centering
	\includegraphics[width=0.48\textwidth]{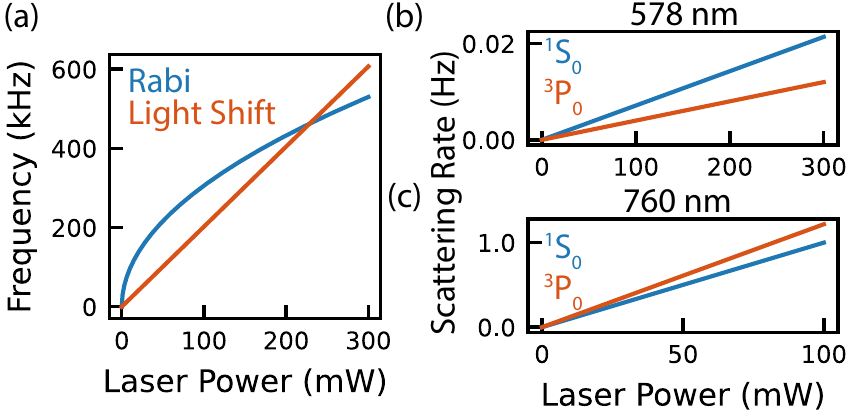}
	\caption{
        \textbf{Off-resonant light shifts and scatter from clock
        pulses and tweezers.} (a) The Rabi frequency $\Omega_c$ and off-resonant
        differential light shift of the clock transition versus the clock laser power for a
        global readout beam (20 $\mu$m beam waist).  (b) \& (c) The off-resonant
        scattering rate of ground (${}^1\t{S}_0$) and metastable (${}^3\t{P}_0$)
        states versus the clock laser power (b) and tweezer power
        (c).}
        \label{FigClockShiftScatter}
\end{figure}

\section{Off-resonant light shifts and scatter from clock pulses and tweezers}\label{ClockShiftScatter}
Beyond magnetic field stability, another important consideration
for the fundamental limitation on coherent evolution is off-resonant dressing
and scattering of the atoms from both the optical pulses and the tweezers. We
focus on the clock pulse in particular since the proposed use case is novel,
while this effect from the Rydberg pulse is universal and similar to other
recent work~\cite{Madjarov2020, Ma2021}.

When an atom is placed into a laser beam, the electrical field
$\mathbf{E}$ causes the atomic dipole moment $\mathbf{p}$ to oscillate at the
driving frequency according to $\mathbf{p} = \alpha \mathbf{E}$, where $\alpha$
is defined as the polarizability of the atom. The real part of the
polarizability introduces the light shift and the imaginary part gives the
scattering of the photons. The calculation of the atom's polarizability gives us
the evaluation of both the light shift and scattering rate, which we consider
for both the clock pulse and tweezers.

As we discussed in the previous section, a Rabi frequency
$\Omega_c = 2\pi \times 200$ kHz for the clock transition is a realistic value
with reasonable polarization purity, temperature and magnetic field stability
requirement. On the other hand, as we increase the Rabi frequency it is also
accompany with the sharp increasing of the clock pulse intensity due to the
relation $\Omega_c \propto \sqrt{P_c}$, where $P_c$ is power of the clock laser.
Since the light shift is proportional to the intensity, the increasing of the
Rabi frequency will also introduce a significant differential light shift
between the ground and clock states that comes from the off-resonant coupling of
the clock laser to all the other transitions. For the tweezers, this
differential light shift is fully canceled under the given clock-magic
wavelength~\cite{Ye2008}.

In Figs.~\ref{FigClockShiftScatter}~(a), we calculate both the
Rabi frequency and the differential light shift introduced by the clock laser
under various laser power for an assumed beam waist radius of $w =
20\,\t{$\mu$m}$. (We focus on global pulses here, but will return to
tightly-focused pulses in Appendix~\ref{AxialClockBeam}.) This suggests the
off-resonant light shift is comparable to the Rabi frequency under our typical
experiment condition, and thus laser intensity noise can be converted into a
noticeable noise of laser detuning $\Delta$. To evaluate the effect of this
power fluctuation, we consider a simple two-level system, where a $\pi$-pulse of
$\Omega_c = 2\pi \times 200\,\t{kHz}$ indicates a square pulse length of $\tau =
2.5\,\t{$\mu$s}$, which corresponds to a window function $W(\delta) =
(\sin(\delta \tau / 2) / (\delta \tau / 2))^2$ that filters out all the noise
with a frequency significant higher than $1 / \tau$, where $\delta$ is the
frequency of the noise. Within this noise bandwidth, a stability better than
$1\%$ is trivial for an active power stabilization setup. For the atom's
transition under this effective frequency noise, a shot-to-shot population
fluctuation of the excited state is $(\Delta / \Omega_c)^2 < (0.01)^2$ where the
differential off-resonant light shift is significantly smaller than the Rabi
frequency. Thus the effect of this differential off-resonant light shifts is
negligible under our usual conditions, but would become significant as $\Omega_c
/ 2\pi$ approaches the MHz scale. On the other hand, a laser pointing error of
$1.4\,\t{$\mu$m}$ will also cause a $1\%$ change of a laser intensity for a beam
waist around $20\,\t{$\mu$m}$. This pointing error noise can either be removed
by the occasionally checking the Rabi frequency during the experiment or by
adding the active position feedback to the mirrors.

Beyond the effect of the differential light shift, the lifetime
and the coherent time of the atom will be limited by the off-resonance
scattering from both the clock pulse and tweezers.
Figs.~\ref{FigClockShiftScatter}~(b) shows the calculation of the scattering
rate on the relevant states of the atoms. The upper plot shows the calculated
scattering rate from the clock laser to the ground and clock state. Under
typical clock pulse intensity, we can find this off-resonance scattering rate is
negligible compared to the $\Omega_c$. Figure~\ref{FigClockShiftScatter}(c) gives the scattering rate of the tweezers
at the clock-magic wavelength, which indicates a more than 1 second lifetime for
both the ground and metastable state atoms under the typical power of the
tweezers. This off-resonance scattering is also negligible for a
$2.5\,\t{$\mu$s}$ clock pulse, and we again emphasize that clock pulses are the
slowest operation in our proposed architecture.

\section{Phase noise analysis}\label{PhaseNoise}
Here we are interested in the dephasing effect of laser phase noise on Rabi
oscillations~\cite{deLeseleuc2018} occurring within the ground-clock manifold.
To analyze this effect under realistic conditions and demonstrate the robustness
of our scheme, we characterize the phase noise from one of our own lasers, tuned
to the  $\ket{\downarrow_\t{g}} \leftrightarrow \ket{\uparrow_\t{m}}$ clock
transition discussed in the main text, and use the measured data in a simulated
drive of the four-state ground-clock manifold following the procedure described
in Appendix~\ref{MultilevelSim}.

\begin{figure}[t!]
	\centering
	\includegraphics[width=0.48\textwidth]{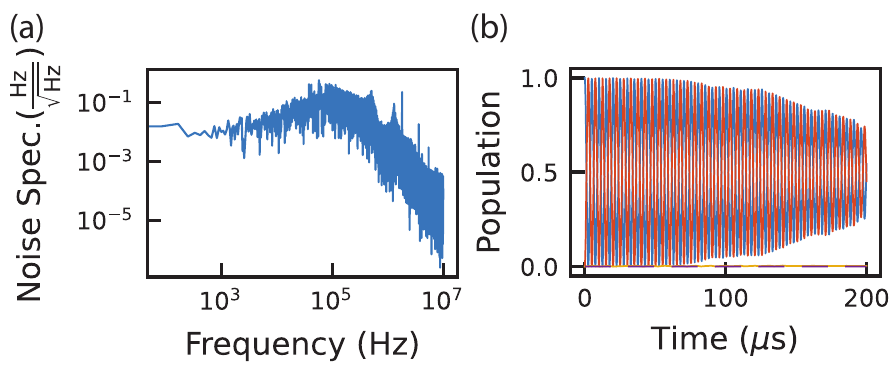}
	\caption{
        \textbf{Phase noise effects from the clock laser.} (a) Measured
        frequency noise spectrum from our cavity reflection without correcting
        for cavity roll-off. This is an approximation of the transmitted signal
        through the cavity that filters phase noise above the cavity bandwidth
        ($\approx 5\,\t{kHz}$). (b) We simulate the ground-clock manifold in the
        presence of a drive with a time-dependent frequency as described in
        Section~\ref{sec: Clock} and Appendix~\ref{MultilevelSim}. This drive
        is measured data, which was used to generate the noise spectrum in (a).
        We find that high-contrast oscillations can be sustained for $\gtrsim
        20$ cycles before dephasing causes decay.
        \label{FigureClockPhaseNoise}
    }
\end{figure}

First, we describe the procedure to characterize the phase noise in the laser.
Our ``clock'' laser ($\lambda = 578\,\t{nm}$) is generated from the second
harmonic of an infrared ``master'' laser at $\lambda_\t{IR} = 1156\,\t{nm}$,
which is locked via the Pound-Drever-Hall (PDH) technique to an ultra-stable
cavity system produced by Stable Laser Systems. We then use the slope of the
in-loop PDH error signal from light reflected from the cavity to obtain the
locked laser's frequency as a function of time and hence compute Allan deviation
and the power spectral density (PSD) of this signal. The measured cavity
response is limited by its linewidth $\nu_c \approx 5\,\t{kHz}$, which gives
significant attenuation of the signal near the frequency band of interest at
$\approx 100\,\t{kHz}$. We could correct for this effect by including a ``cavity
roll-off factor''~\cite{Tarallo2009} to accurately portray the phase noise on
our laser, but in this work we consider using the transmitted light through the
cavity to filter this phase noise~\cite{Levine2018}. Hence, the phase noise of
the transmitted light is accurately represented by our measurement of the
reflected light directly, without including the cavity roll-off factor. We
believe this approach will make our analysis more generally applicable. With
this procedure we calculate the Allan deviation of the measured signal to be
$\sigma \lesssim 2 \times 10^{-15}$ at a $\tau = 1\,\t{s}$ averaging time and
estimate the linewidth of the laser to be $\Delta\nu \approx 2\,\t{Hz}$ from the
PSD, shown in Fig.~\ref{FigureClockPhaseNoise}(a), using the $\beta$-separation
line method~\cite{Domenico2010}.

The phase noise data was then used to generate a realistic, time-dependent drive
to a simulated four-level ground-clock manifold. This is accomplished by taking
a sum over Fourier components that are weighted by the calculated PSD with
random phase shifts sampled from a uniform distribution. When applied in
simulation following the description given in Appendix~\ref{MultilevelSim}, we
find that high-contrast Rabi oscillations can be sustained over more than 20
cycles with this drive, as shown in Fig.~\ref{FigureClockPhaseNoise}(b).

\section{Finite temperature modeling}\label{Thermal}
We now incorporate finite-temperature effects in our analysis of single-atom
dynamics. In an optical tweezer, a single atom at non-zero temperature is
delocalized over lengths comparable to the wavelength of the laser; hence we
must include a position-dependent motional phase factor $\exp(i \mathbf{k} \cdot
\mathbf{x})$ into the drive $\Omega$, where $\mathbf{k}$ is the wavevector of
the driving laser. $\eta=kx_0$ is the Lamb-Dicke parameter (see the main text).
For simplicity, we approximate the tweezer with a one-dimensional harmonic
trapping potential~\cite{Kale2020, Jenkins2021} and write the motional phase
factor as $\exp(i \eta (\hat{a} + \hat{a}^\dag))$, where $\hat{a}$ and
$\hat{a}^\dag$ are ladder operators operating on the Fock basis $\{\ket{n}\}$
corresponding to the usual harmonic oscillator states. For brevity, we denote
the motional phase factor and its adjoint as $\hat{\xi} = \exp(i \eta (\hat{a} +
\hat{a}^\dag))$ and $\hat{\xi}^\dag = \exp(-i \eta (\hat{a} + \hat{a}^\dag))$.
For the case of a ``magic'' wavelength trap (where the atomic ground and excited
states experience the same trap frequency), the Hamiltonian of the system
is~\cite{Kale2020, Jenkins2021}
\begin{equation}\label{eq:Hamitonian_FiniteT}
    \begin{split}
	\hat{H}
        &= \hbar\sum_{g,e} \frac{\Omega}{2}
            \Big(
                g_g^e(\chi, q) e^{i (\omega - \omega_0 - \tilde{\omega}_g^e)}
                    \hat{\xi} \otimes \ketbra{e}{g}
                + \t{H.c.}
            \Big) 
        \\&\phantom{=}
        + \hbar \omega_t \left(
            \hat{a}^\dag \hat{a} + \frac{1}{2}
        \right) \otimes \mathbb{\hat{I}}
    \end{split}        
\end{equation}
where $\ket{g} \in \left\{\ket{\downarrow_\t{g}},\,
\ket{\uparrow_\t{g}}\right\}$ and $\ket{e} \in \left\{\ket{\downarrow_\t{m}},\,
\ket{\uparrow_\t{m}}\right\}$.  $\mathbb{\hat{I}}$ is the $4 \times 4$ identity
operator for the four-level ground-clock manifold. $\Omega g_g^e(\chi, q)$ is
the driving term which includes both the effects of polarization impurity and
Clebsch-Gordan weighting factor (see Appendix~\ref{MultilevelSim}).

For our purposes, we consider $\Omega \gg \omega_r$ for a high-fidelity state
transfer. The higher-order terms of $\hat{\xi}$ are also no longer strongly
suppressed and couple a single motional state to many other excited motional
states at the same time. To simplify calculations we rewrite the basis states of
the combined atom-Fock Hilbert space as $\ket{g, n} = \ket{g} \otimes \ket{n}$
and $\ket{e, \xi(n)} = \ket{e} \otimes \hat{\xi} \ket{n}$. We then rewrite the
Hamiltonian by inserting the identity resolved in this basis to the left and
right,
\begin{equation}\label{eq:Newbasis_FiniteT}
    \begin{split}
        \hat{H}
            &\rightarrow \left(
                \sum_{n', g', e'} \ketbra{g', n'}{g', n'}
                + \ketbra{e', \xi(n')}{e', \xi(n')}
            \right) \hat{H}
            \\&\phantom{=}
            \times \left(
                \sum_{n, g, e} \ketbra{g, n}{g, n}
                + \ketbra{e, \xi(n)}{e, \xi(n)}
            \right)
    ,\end{split}
\end{equation}
and define a four-level state vector $G_n$ for the $n$-th motional state
\begin{equation}\label{eq:4level_State_vector}
	G_n
        = \Big(
            \ket{\uparrow_\t{m},\xi(n)},
            \ket{\downarrow_\t{m},\xi(n)},
            \ket{\uparrow_\t{g},n},
            \ket{\downarrow_\t{g},n}
        \Big).
\end{equation}
Thus the Hamiltonian can be simplified as:
\begin{equation}\label{eq:Newbasis_FiniteT_Simplified}
    \begin{split}
        \hat{H}
            &\rightarrow \left(\sum_{n'} G_{n'} G_{n'}^\dag\right) \hat{H}
                \left(\sum_{n} G_{n} G_{n}^\dag\right)
            \\
            &= \sum_{n,n'} G_{n'}^\dag \left(G_{n'} \hat{H} G_{n}^\dag\right)
                G_{n},
    \end{split}
\end{equation}
where $G_{n'} \hat{H} G_{n}^\dag$ is a $4 \times 4$ matrix. The Hamiltonian can
then be understood as a $4 \times 4$ matrix under $N^2$ different conditions
that describe the transitions between different motional states. These
individual $4 \times 4$ matrices can then be assembled into a $N \times N$ table
to reduce computer memory usage in numerical computation, where $N$ is the
highest motional state we want to include in the calculation. For our
calculations, we use $N = 100 \gg k_B T / \hbar \omega_r$. For a given
temperature, we use the appropriate Boltzmann distribution to construct an
initial state vector, and numerical simulation is accomplished by the method
described in Appendix~\ref{MultilevelSim}.

\section{Linearly-polarized drives}\label{Linear}
\begin{figure}[t!]
	\centering
	\includegraphics[width=0.48\textwidth]{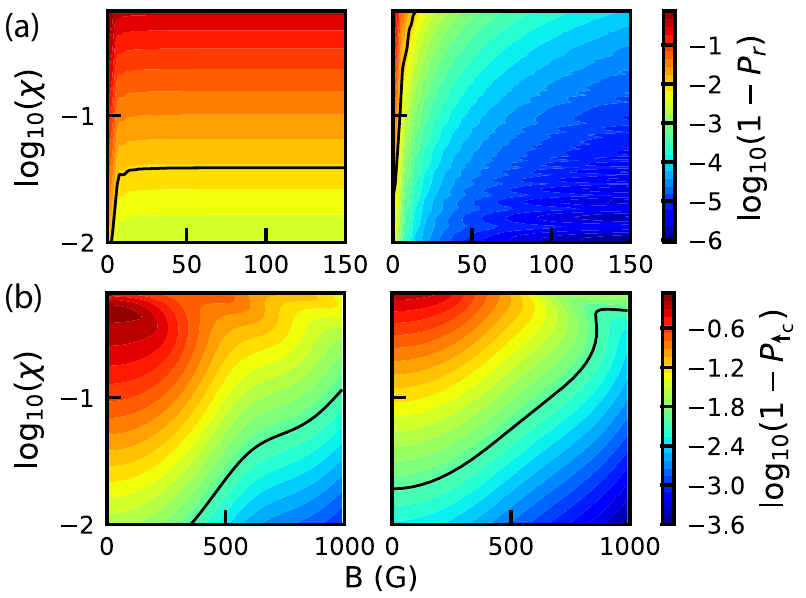}
	\caption{
        \textbf{Comparison of $\pi$-pulse infidelity for linearly ($\pi$) and
        circularly ($\sigma^+$) polarized drives.} (a) Infidelity for the
        clock-Rydberg transition case for the linear (left) and circular (right)
        drive. (b) Infidelity for the ground-clock transition case for the
        linear (left) and circular (right) drive. The black line indicates where
        the infidelity crosses $0.01$.
        \label{FigureClockPolCompare}
    }
\end{figure}

In this analysis, we compare the cases of driving the aforementioned transitions
with linearly-polarized ($\pi$) and circularly-polarized ($\sigma^{+}$) light.
The $\pi$-polarized drives target the ground-clock $\ket{\downarrow_\t{g}}
\leftrightarrow \ket{\downarrow_\t{m}}$ and clock-Rydberg $\ket{\uparrow_\t{m}}
\leftrightarrow \ket{r_\uparrow}$ transitions, giving $\Delta m_F = 0$ as
opposed to $\Delta m_F = +1$ for the $\sigma^{+}$ transitions.
Fig.~\ref{FigureClockPolCompare} shows the $\pi$-pulse infidelities for both
cases, providing a direct comparison between the $\pi$ and $\sigma^{+}$
transitions under various polarization impurities $\chi$ and magnetic field
strengths $B$.

The $\pi$ drives introduce greater sensitivity to polarization impurity,
particularly for the clock-Rydberg case, because a resonant Raman condition
exists between the two nuclear spin states for $\chi > 0$. In contrast, this
condition does not exist for the clock-Rydberg case with $\sigma^+$ drives since
the target state is stretched to maximum $m_F$. Hence, driving the $\pi$
transition with high fidelity requires very low polarization impurity ($\chi <
10^{-2}$) and shows minimal improvement with larger magnetic fields. In contrast,
$\sigma^+$ drives yield significantly greater populations in the target state
$\ket{r}$ while exhibiting a much higher tolerance to impurity. Driving the
four-level ground-clock transition with $\pi$-polarization is also inferior to
$\sigma^+$ with similar reasoning. Fig.~\ref{FigureClockPolCompare}(b)
highlights important distinctions between the drives across lower magnetic
fields. We find that $\pi$-driven clock transitions with $\chi > 10^{-2}$ require
larger magnetic field for the same pulse fidelity compared to the $\sigma^+$
case.

\section{Varying the clock-transition Rabi frequency}\label{ClockRabi}
In the main text, we primarily consider the use of $\Omega / 2 \pi =
200\,\t{kHz}$ for the clock transition Rabi frequency. Here, we vary $\Omega$,
neglecting motion and thermal effects, to identify conditions under which the
nuclear spin splitting will limit the $\pi$-pulse fidelity and Rabi coherence
time. We study the population $P_{\uparrow_\t{m}}$ in $\ket{\uparrow_\t{m}}$
after a $\pi$- and $9 \pi$-pulse from $\ket{\downarrow_\t{g}}$ versus magnetic
field and Rabi frequency with $\chi = 10^{-2}$ (see
Fig.~\ref{FigureClockRabiCompare}). For sufficiently high $\Omega$, the pulse
fidelity after $9 \pi$ is worse than that of $\pi$, which indicates the onset of
non-negligible coupling to the spectator states. Interestingly, we observe
non-monotonic behavior with respect to varying $\Omega$, which we attribute to
resonance effects where the Zeeman shift of the spectator transition (which
depends on $B$) is within the bandwidth of $\Omega$. For higher $\chi$ this
effect would become crippling even for relatively short pulses.

\section{Axial addressing \& driving the clock transition at lower trap frequency}\label{AxialClockBeam}
Besides the global clock pulse, a tightly-focused beam could be
useful for single-qubit, mid-circuit readout. In a tweezer system, this can be
accomplished by overlapping the clock laser together with a tweezer. Since a
tweezer has much weaker confinement on the axial direction compared to the
radial direction, the consideration of the motional states change dramatically.

\begin{figure}[t!]
	\centering
	\includegraphics[width=0.48\textwidth]{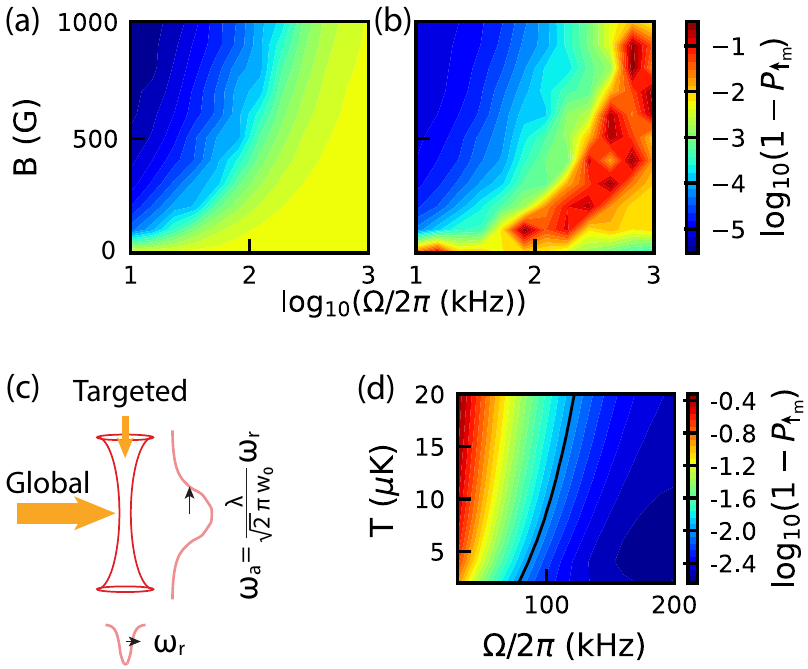}
	\caption{
        \textbf{Driving the clock transition for various Rabi and trap
        frequencies.} (a) $\pi$-pulse and (b) $9\pi$-pulse, initialized in
        $\ket{\downarrow_\t{g}}$, under various magnetic field ($B$) and Rabi
        frequency ($\Omega$) with $\chi = 10^{-2}$. We neglect motion and
        thermal effects. The color scale is the population in
        $\ket{\uparrow_\t{m}}$, $P_{\uparrow_\t{m}}$. (b) shows that some
        non-monotonic behavior develops at high $\Omega$ due to increased
        coupling to the spectator states. (c) Diagram of
        targeted and global pulses and the corresponding relevant tweezer
        trapping axes. (d) The infidelity due to the finite temperature effects
        for a clock beams on the axial direction.}
	\label{FigureClockRabiCompare}
\end{figure}

Similar to the calculation for the radial direction in
Appendix~\ref{Thermal} and assuming a Gaussian beam profile with fixed
waist-Rayleigh length relationship, we assume an axial trap frequency in the
tweezer of $\omega_z = 2\pi\times 17\,\t{kHz}$ [corresponding to the same
tweezer parameter in the main text; see Fig.~\ref{FigureClockRabiCompare}(c)].
Under this lower trap frequency, the simulation shown in
Fig.~\ref{FigureClockRabiCompare}(d) indicates a significantly higher fidelity
for most cases when compared with the radial direction. This result can be
understood by considering that an atomic transition is mostly affected by phase
noise around the Rabi frequency. The atomic sloshing motion can introduce an
effective phase noise around the trap frequency, which is particularly
deleterious when the trap frequency is still comparable to the Rabi frequency.
For the case of axial addressing, since the Rabi frequency is much higher than
the trap frequency, the atom is nearly static during the clock pulse time.

However, other technical problems arise when we apply the axial
driving beam, which also pertain to focused Raman-based single-qubit gates and focused Rydberg-mediated two-qubit gates [see Fig.~\ref{OMG}(b)]. One issue is the motional stability of the tweezers and clock laser
beams, which requires a relative displacement smaller than $0.1\,\t{$\mu$m}$ for
an intensity fluctuation smaller than $1\%$ assuming a waist of $800\,\t{nm}$.
This issue can be solved by adding a flat-top beam shaper, which gives a
homogeneous clock laser intensity within the beam diameter. Another potential
problem is the crosstalk between the target atom and its neighbors. However,
considering a tweezer spacing of $2 ~ (3)\,\t{$\mu$m}$ and an addressing beam
waist radius of $800\,\t{nm}$, the laser Rabi crosstalk of the neighboring atoms
is $\approx10^{-3}$ ($\approx10^{-6}$), which suggests that the operations in
our architecture can exceed the 0.99 fidelity level.

Another insidious technical issue for tightly focused beams of highly constrained polarization (especially if it is circular such as the clock and Rydberg beams) is the need to maintain this polarization for all sites to be addressed by the tightly focused beams. However, this may be accomplished by carefully designing the optical system to put polarizers in the appropriate plane, perhaps combined with the use of metallic mirrors rather than dielectric mirrors. Additionally, the Pockels electro-optic effect could be used to adjust the polarization in a calibrated map via a polarimeter. Although we leave a careful study of this effect for future work, we note that trapped ion systems have been engineering solutions to such problems for over a decade~\cite{Leibfried2003}.

\section{Technical limitations for the Rydberg transition}\label{RydbergNoise}
As stated above, we believe that the technical limitations of driving ground-Rydberg transitions for use in Rydberg-mediated entanglement are thoroughly described elsewhere~\cite{Levine2018,deLeseleuc2018,Madjarov2020}, but we briefly consider them in the context of our architecture. A recurring theme is the disparate timescales between the clock drives and Rydberg drives ($\Omega_\text{R}>10\Omega_\text{c}$), rendering the Rydberg drives less sensitive to several technical limitations.

\subsection{Laser frequency noise}
Closed-loop frequency stabilization systems introduce noise peaks, called ``servo bumps", that typically span $\approx100$ kHz to $\approx1$ MHz. This frequency noise gives rise to a $\Delta(t)\sigma_z$ term that must be considered in addition to the $\Omega\sigma_x$ Rabi drive term, and is well known to have particularly deleterious effects when its characteristic timescale $\tau$ matches $1/\Omega$~\cite{Levine2018,deLeseleuc2018,Madjarov2020}. As described above, we consider $\Omega_\text{c}=2\pi\times200$ kHz for the clock transition and $\Omega_\text{R}=2\pi\times6$ MHz for the Rydberg transition. Therefore, the clock drive is substantially more sensitive to laser frequency noise than the Rydberg drive. Indeed, this setting for the Rydberg transition was recently used in a nearly identical system with $^{88}$Sr~\cite{Madjarov2020}, showing long-time Rabi coherence with contrast exceeding 0.99.

\subsection{Motional and trapping effects}
Unlike the clock transition for which the differential polarizability is zero at 759 nm, there is a significant differential polarizability at this wavelength for the Rydberg transition~\cite{Madjarov2020}. It is common to blink the traps off during Rydberg pulses. (The intended Rydberg-based gates will not leave population in the Rydberg state after the pulse.) With the atom in free flight, its motion gives rise to random Doppler shifts given by $\Delta\omega=2\pi/\lambda\sqrt{k_\text{B}T/m}$, where $\lambda$ is the optical wavelength, $T$ is the temperature, $k_\text{B}$ is Boltzmann's constant, and $m$ is the mass. Assuming a temperature of 500 nK at a trap depth of 5 $\mu$K (adiabatically ramping down from a temperature of 5 $\mu$K in a 500 $\mu$K-deep trap), $\Delta\omega\approx2\pi\times16$ kHz. This effect is negligible compared to $\Omega_\text{R}=2\pi\times6$ MHz.

Alternatively, one could leave the tweezer traps on during the pulses. As discussed above, the trap frequencies in a 500 $\mu$K-deep trap are at most $\approx$70 kHz. While this is comparable to $\Omega_\text{c}$, it is much smaller than $\Omega_\text{R}$. Moreover, it is common to ramp the trap depth down by a factor of 100 bringing us to $U=5$ $\mu$K $\approx100$ kHz, for which this frequency is $\approx$7 kHz. In this setting, $\Omega_\text{R}\gg U$. In this limit, the dominant effect from the trap is the random differential light shift due to the deviation of the atomic position from the trap bottom. Similar to the free-space case, this corresponds to $\Delta\omega\approx2\pi\times15$ kHz when assuming an atomic temperature of 500 nK in the 5 $\mu$K-deep trap, and assuming a relative polarizability of $\alpha_\text{R}/\alpha_\text{c}\approx-0.5$~\cite{Wilson2019,Madjarov2020}. The recent work with Sr~\cite{Madjarov2020} also studied the case with the traps on, finding minimal difference versus blinking them off.

\subsection{Rydberg state lifetime}
Here again, large $\Omega_\text{R}$ helps to mitigate the effects of decay from the Rydberg state, which again has been discussed in detail~\cite{deLeseleuc2018}. We consider the use of a Rydberg state with $n^*\approx55$, for which we anticipate a lifetime of $\tau\approx100$ $\mu$s. The $2\pi$-pulse of our Rydberg gates is $\tau_{2\pi}=2\pi/\Omega_\text{R}=167$ ns, which suggests that pulses with fidelity up to $\gtrsim$0.999 are possible when integrating over the population of the Rydberg state during the pulse. Working at cryogenic temperatures can further improve the Rydberg state lifetimes.

\subsection{DC Stark and Zeeman effects}
Finally, we consider DC drifts in the resonance frequency of the Rydberg transition originating from magnetic and electric field instability. The former is already considered in Appendix~\ref{MultilevelSim} where we assumed a 1 mG field instability and showed negligible effects, again owing to the large separation between $\Omega_\text{R}=2\pi\times6$ MHz and $\Delta_B=2\pi\times1.9~\text{MHz/G}\cdot10^{-3}~\text{G}=2\pi\times1.9$ kHz. We do not anticipate DC Stark shifts that are significantly different than those of other atomic species with comparable $n^*$~\cite{Levine2019,Madjarov2020}, and thus we do not anticipate limitations even well beyond the 0.99 level due to this effect for $n^*\approx55$, albeit perhaps requiring interleaved lineshape measurements or active atomic locking~\cite{Madjarov2020,Choi2021}. In-vacuum electrode systems~\cite{Covey2018a,deLeseleuc2018,Wilson2019} can further suppress the effect of electric field transients, and so can operation at cryogenic temperatures.

\bibliography{library}

\end{document}